\documentclass[twocolumn,superscriptaddress,pra]{revtex4-2}
 
\usepackage{amsmath, amssymb, amsthm}
\usepackage{mathtools}
\usepackage{xurl}
\usepackage{float}
\usepackage{graphicx} 
\usepackage{tikz}
\usepackage{caption}
\usepackage{pgfplots}
\pgfplotsset{compat=1.18}
\usetikzlibrary{shapes.geometric, arrows.meta, positioning, calc}
\usetikzlibrary{patterns,decorations.pathreplacing,calligraphy}
\usepackage{booktabs}
\usepackage{hyperref}
\hypersetup{
    colorlinks=true,
    linkcolor=blue,
    citecolor=blue,
    urlcolor=blue
}
\usepackage{algorithm}
\usepackage{algorithmic}
 
\allowdisplaybreaks

\begin{document}

\title{Open vs. Sealed: Auction Format Choice for Maximal Extractable Value}
 
\author{Aleksei Adadurov}
\author{Sergey Barseghyan}
\author{Anton Chtepine}
\author{Antero Eloranta}
\author{Andrei Sebyakin}
\author{Arsenii Valitov}
\affiliation{nuconstruct \\ \textnormal{\texttt{team@nuconstruct.xyz}}}
 
\date{\today}
 
\begin{abstract}
We study optimal auction design for Maximum Extractable Value (MEV) auction markets on Ethereum. Using a dataset of 2.2 million transactions across three major orderflow providers, we establish three empirical regularities: extracted values follow a log-normal distribution with extreme right-tail concentration, competition intensity varies substantially across MEV types, and the standard Revenue Equivalence Theorem breaks down due to affiliation among searchers' valuations. We model this affiliation through a Gaussian common factor, deriving equilibrium bidding strategies and expected revenues for five auction formats, first-price sealed-bid, second-price sealed-bid, English, Dutch, and all-pay, across a fine grid of bidder counts $n$ and affiliation parameters $\rho$. Our simulations confirm the Milgrom-Weber linkage principle: English and second-price sealed-bid auctions strictly dominate Dutch and first-price sealed-bid formats for any $\rho > 0$, with a linkage gap of 14-28\% at moderate affiliation ($\rho=0.5$) and up to 30\% for small bidder counts. Applied to observed bribe totals, this gap corresponds to \$10-18 million in foregone revenue over the sample period. We also document a novel non-monotonicity: at large $n$ and high $\rho$, revenue peaks in the interior of the affiliation parameter space and declines thereafter, as near-perfect correlation collapses the order-statistic spread that drives competitive payments. 
\end{abstract}


\maketitle

\newpage

\section{Introduction}

Block builders auction tens of millions of dollars worth of Maximum Extractable Value opportunities to competing searchers each year in the Ethereum ecosystem. These auctions determine how the surplus from transaction ordering is divided between searchers, who identify and execute profitable strategies, and orderflow providers, who control access to transaction flow. Despite the economic magnitude of these markets, relatively little attention has been paid to the question of mechanism optimality: which auction format should a block builder use to maximize their own revenue?

This question is non-trivial for several reasons. First, standard auction theory treats valuation distributions as a background parameter, but the shape of that distribution turns out to matter enormously here. Our empirical analysis shows that MEV valuations follow a log-normal distribution, where the top 1\% of transactions generate 68\% of all revenue. When value is this concentrated in the right tail, mechanism choice for high-value transactions dominates overall revenue performance, and formats that differ only in their handling of typical transactions may diverge sharply on the rare events that matter most.

Second, competition is highly heterogeneous across MEV types. Sandwich attacks, which exploit publicly observable mempool information, attract intense competition and exhibit average bribe percentages above 95\% indicating that searchers bid away nearly all of their surplus. By contrast, naked arbitrage and liquidations attract fewer searchers, with correspondingly lower bribe ratios. A single auction format optimized for one competitive regime will generally be suboptimal in another, suggesting that mechanism design for MEV markets should be segmented rather than uniform.

Third, standard revenue equivalence arguments break down in this environment. The Revenue Equivalence Theorem establishes that all standard auction formats yield identical expected revenue when bidders hold independent private values, are symmetric and risk-neutral, and the object is allocated to the highest-value bidder. But MEV valuations are plausibly affiliated: if market conditions make a cross-DEX arbitrage opportunity valuable to one searcher, they tend to make it valuable to others, since all searchers observe the same on-chain state and off-chain price signals. Under affiliation, the linkage principle of Milgrom and Weber (1982) \cite{milgrom1982} predicts a strict revenue ranking: formats with truthful bidding (English and second-price sealed-bid) outperform formats with strategic bid shading (first-price sealed-bid and Dutch).

We address these questions through a combination of empirical analysis, equilibrium theory, and numerical simulation. Using a dataset of 2.2 million transactions, documenting MEV extraction across Blink, Merkle, and MEV Blocker, we first establish the empirical regularities that motivate our theoretical framework. We then derive equilibrium bidding strategies and expected revenues for five auction formats under log-normal valuations, using the Gaussian copula construction to model affiliation. Finally, we simulate revenue outcomes across a fine grid of bidder counts and affiliation levels and compare counterfactual revenues for each format using the empirical transaction data.

Our main findings are as follows. Under an independent private values setting, all five formats satisfy revenue equivalence, consistent with theory and confirmed by our Monte Carlo verification. Under affiliation, a clear ranking emerges: English and SPSB auctions dominate Dutch and FPSB auctions, which in turn dominate the all-pay format. At the baseline affiliation level of $\rho = 0.5$ with $n = 5$ bidders, English and SPSB auctions yield revenues 17\% higher than Dutch and FPSB formats, and the all-pay mechanism falls 41\% below the FPSB baseline. Applied to the empirical transaction data, these differences translate to substantial sums: switching from a first-price to second-price or English format would have increased auctioneers' revenues by between 3.3\% and 32.0\%, though the magnitude is sensitive to the assumed affiliation parameter and the number of bidders. We also document a novel non-monotonicity: at high affiliation and large bidder counts, revenue peaks in the interior of the affiliation parameter space and then declines as near-perfect correlation collapses the order-statistic spread that drives competitive payments. 

The remainder of the paper proceeds as follows. Section 2 reviews relevant auction theory and the existing literature on MEV auction mechanisms. Section 3 describes the data and documents the empirical regularities. Section 4 develops the theoretical framework and derives equilibrium characterizations for each auction format. Section 5 reports the numerical results and robustness checks. Section 6 concludes the paper.

\section{Auction Theory and Related Work}

\subsection{Auction Theory and Mechanism Design}

Auctions serve as a fundamental mechanism for price discovery across diverse markets, from government bonds and online advertising placements to construction contracts and airport landing slots. By eliciting competitive bids from participants with private valuations, auctions reveal willingness-to-pay and facilitate efficient allocation of scarce resources \cite{mcafee1987}. Different auction formats such as English (or, ascending), Dutch (or, descending), first-price sealed-bid, second-price sealed-bid (or, Vickrey), and all-pay each possess distinct properties that make them suited to particular market contexts \cite{klemperer2004}. The choice of auction design influences bidding behavior, revenue outcomes, allocative efficiency, and fairness, depending on factors such as the number of participants, information asymmetries, and the nature of the good being traded \cite{milgrom2004}.

Each auction format presents a distinct tradeoff in terms of efficiency, revenue generation, and strategic complexity. English auctions promote price discovery through transparent, sequential bidding, allowing participants to observe competitors' valuations and adjust strategies accordingly; however, this transparency can facilitate collusion and may disadvantage bidders with slower information processing or execution capabilities \cite{klemperer2004, mcafee1987}. Dutch auctions offer speed and simplicity, concluding as soon as any bidder accepts the current price, but they provide limited price discovery and may result in suboptimal outcomes when bidders face uncertainty about asset value or competitor valuations \cite{krishna2009}. Vickrey auctions are theoretically optimal in encouraging truthful bidding, since participants pay the second-highest bid rather than their own, thereby simplifying strategy and promoting allocative efficiency \cite{vickrey1961, myerson1981}; however, they are vulnerable to manipulation through rigged bidding, require strong trust in the auctioneer's execution, and are therefore rarely used in high-stakes settings \cite{ausubel2006}.

\subsection{Comparison of Auction Formats}

The choice of auction format fundamentally shapes bidding behavior, revenue outcomes, and allocative efficiency. Here, we compare different auction types such as English, Dutch, first-price sealed-bid, second-price sealed-bid, and all-pay auctions, analyzing the mechanism of each alongside its strategic properties and implications for both auctioneers and bidders.

\subsubsection{English}

In an English auction, the auctioneer announces successively higher prices and bidders publicly signal their willingness to continue at each level. The auction concludes when only one active bidder remains, who wins at the final price. Under the independent private values (IPV) framework, where each bidder's valuation is drawn independently from a common distribution and is known only to themselves, the dominant strategy is to remain active until the price reaches one's true valuation and then drop out \cite{krishna2009}. Dropping out early risks losing a profitable win, while remaining active beyond one's valuation guarantees a loss if successful \cite{milgrom1982}. Truthful bidding thus emerges naturally, without requiring bidders to reason about competitors' valuations \cite{krishna2009}.

For the auctioneer, the winner pays approximately the second-highest valuation, making expected revenue equal to the expected second-order statistic of the valuation distribution \cite{myerson1981, riley1981}. The sequential, public nature of bidding also provides valuable price discovery, as bidders can observe dropout behavior and update their beliefs about aggregate demand, which tends to enhance allocative efficiency \cite{milgrom1982}. However, this same transparency creates collusion risk: bidders can coordinate through observed dropout patterns, potentially suppressing prices \cite{mcafee1987, klemperer2004}. Slower or less sophisticated participants may also be disadvantaged by the pace of real-time competition \cite{klemperer2004}.

\subsubsection{Dutch}

A Dutch auction begins at a high price, which the auctioneer lowers continuously until a bidder accepts, winning the item at that price. Although procedurally opposite to the English format, the Dutch auction is strategically equivalent to the sealed-bid first-price auction under the IPV setting. In both cases, bidders face the same fundamental trade-off: accepting or submitting a lower price increases the potential surplus from winning, but reduces the probability of winning \cite{krishna2009}.

The strategic implication is that rational bidders shade their bids below their true valuations \cite{myerson1981}. For the auctioneer, this bid-shading means the Dutch format does not induce truthful revelation and, absent strong competition, may yield lower realized prices than English auctions \cite{klemperer2004}. For bidders, the Dutch format demands more sophisticated reasoning than the English format, since there is no sequential information to guide decisions \cite{milgrom1982}. The speed of the Dutch clock can also disadvantage bidders who need more time to deliberate, with the risk that hesitation causes them to lose to a faster but lower-valuing competitor \cite{klemperer2004}.

\subsubsection{First-Price Sealed-Bid}

In a first-price sealed-bid (FPSB) auction, all bidders simultaneously submit private offers without observing one another's bids. The highest bidder wins and pays their submitted amount. As in the Dutch auction, the optimal strategy involves shading one's bid below true valuation, balancing the gain from a lower price against the risk of being outbid \cite{myerson1981}. Competition intensifies this trade-off: as the number of bidders increases, bid-shading diminishes because the probability of losing to a higher bid rises \cite{krishna2009}.

For the auctioneer, the FPSB format reduces collusion risk relative to the English auction, since bidders cannot observe and react to one another's behavior \cite{klemperer2004}. However, the auctioneer cannot guarantee truthful revelation, and expected revenue depends critically on the distribution of valuations and the number of participants. A further complication arises in common value settings, where the item's true value is uncertain and correlated across bidders: winners systematically overbid because winning itself signals that one's estimate exceeded all others', suggesting it was likely inflated \cite{milgrom1982, bulow2002}. This winner's curse is a particular hazard for less experienced bidders \cite{bulow2002}. On the other hand, when bidders are risk-averse, FPSB auctions can generate higher expected revenue than English or Vickrey auctions, since bidders bid more aggressively to secure the item with certainty \cite{maskin1984}.

\subsubsection{Second-Price Sealed-Bid (Vickrey)}

Vickrey auctions modify the FPSB format by requiring the winner to pay the second-highest bid rather than their own \cite{vickrey1961}. This payment rule has a profound strategic consequence: truthful bidding, submitting one's true valuation, is a weakly dominant strategy \cite{myerson1981}. Overbidding cannot improve the price paid (which is set by others), and only risks winning at a loss \cite{vickrey1961}. Underbidding only risks forfeiting a profitable win when the second-highest bid falls between the shaded bid and true valuation \cite{myerson1981}. Bidders therefore need not engage in strategic reasoning about competitors at all \cite{krishna2009}.

For the auctioneer, expected revenue equals the expected second-highest valuation, which the Revenue Equivalence Theorem establishes as identical to English and FPSB auction revenue under standard conditions \cite{riley1981, myerson1981}. The dominant strategy property also promotes allocative efficiency, since the highest-valuing bidder always wins \cite{krishna2009}. Despite these theoretical strengths, the Vickrey format faces practical difficulties. It is vulnerable to shill bidding, in which the auctioneer or a colluding party inflates the second-highest bid to raise the price paid \cite{ausubel2006}. Winning bidders may also perceive the outcome as unfair or suspect manipulation when the price they pay appears suspiciously close to their own bid \cite{ausubel2006}. These concerns have limited the adoption of Vickrey auctions in practice, outside specialized contexts such as online advertising and blockchain-based mechanisms \cite{ausubel2006}.

\subsubsection{All-Pay}

All-pay auctions depart sharply from standard formats: every bidder pays their submitted bid regardless of outcome, and the highest bidder wins the item. This structure models competitive situations where resources are irreversibly committed during the contest, such as lobbying, patent races, or litigation, where only one party prevails but all incur costs \cite{baye1996}.

The strategic implication for bidders is substantial bid compression. Since losing bidders still pay, the expected cost of aggressive bidding is much higher than in other formats, leading bidders to shade their bids more dramatically than in FPSB auctions \cite{krishna1997}. In common value settings, the all-pay structure compounds the winner's curse, since bidders must pay regardless of outcome and therefore bear the full cost of any valuation error \cite{milgrom1982}. These features make all-pay auctions appropriate only in contexts where irreversible commitment is an intrinsic feature of the competition rather than a design choice \cite{baye1996}.

\subsection{Auctions in Blockchain Ecosystems}

In blockchain ecosystems, auctions have emerged as a foundational mechanism for allocating transaction inclusion rights within blocks. In decentralized networks like Ethereum, block builders or miners effectively run continuous auctions where users compete to have their transactions included, often by attaching higher fees. These crypto transaction auctions manifest in various formats depending on the specific mechanism: MEV searchers participate in sealed-bid first-price auctions when competing to capture arbitrage and liquidation opportunities; block builders engage in open/English auctions to win the right to propose blocks to validators under MEV-Boost; Arbitrum’s Timeboost employs a sealed-bid second-price auction for express lane transaction inclusion rights; and protocols like Compound have a liquidation mechanism with time-decaying discount factors.

These auctions are distinctive in their high-frequency, real-time nature and strategic complexity. In particular, MEV searchers operate as sophisticated algorithmic traders who scan for profitable opportunities, such as arbitrage across decentralized exchanges, liquidations of undercollateralized loans, or sandwich attacks around large trades, and submit bundles of transactions with strategic bids to capture value. Unlike traditional auctions, crypto transaction auctions are executed under strict time constraints, limited observability (due to private mempools and encrypted order flow as well as timing), and evolving incentive structures shaped by protocol-level changes. The interplay between users, validators, searchers, and order-flow intermediaries creates a dynamic competitive environment in which bidding strategies and auction formats significantly influence network efficiency, fairness, and revenue distribution.

\subsection{Auction Design for Orderflow in MEV Contexts}

The Maximal Extractable Value (MEV) in blockchain ecosystems has led to the development of specialized auction mechanisms designed to allocate transaction ordering rights and capture value that would otherwise be extracted by miners, validators, or sophisticated searchers. Unlike traditional auctions, MEV-related auctions operate in high-frequency, time-constrained environments where millisecond-level latency, information asymmetries, and complex strategic interactions fundamentally reshape bidding behavior and auction outcomes.

Following Ethereum's transition to Proof-Of-Stake, Proposer-Builder Separation (PBS) emerged as the dominant mechanism for block construction, with MEV-Boost serving as its primary implementation. MEV-Boost operates as an open ascending auction, where block builders compete to have their blocks selected by validators \cite{pai2023, wu2024}. In this auction, builders submit open bids alongside block headers to relays, which act as trusted intermediaries. The proposer ultimately selects the highest-bidding builder, who pays their bid to the validator in exchange for block inclusion rights.

Wu et al. \cite{wu2024} provide a game-theoretic analysis of MEV-Boost auctions, examining how builders' access to MEV opportunities and network connectivity influence strategic bidding. Their simulation-based approach reveals that the interplay between technical infrastructure advantages, particularly low-latency connections to relays, and exclusive access to profitable transaction bundles drives market concentration. Empirical evidence from EigenPhi's analysis of MEV-Boost bidding data \cite{eigenphi2024} confirms these theoretical predictions, documenting that dominant builders engage in aggressive late-bidding strategies. This timing, combined with high transaction volumes, creates substantial barriers to entry for smaller builders and reinforces centralization tendencies.

Öz et al. \cite{oz2024} investigate the determinants of auction success in MEV-Boost, finding that builder profitability exhibits significant heterogeneity. While the top three builders capture the majority of blocks, their profit margins vary considerably, ranging from slightly negative to over 5\% of block value. This heterogeneity reflects differences in MEV extraction capabilities, orderflow access, and bidding sophistication rather than pure auction design characteristics.

The role of private orderflow in shaping auction outcomes is one of the central concerns in MEV research. Gupta, Pai, and Resnick \cite{gupta2023} analyze how exclusive access to private transactions confers competitive advantages in PBS auctions. Integrated builder-searchers who receive private orderflow can extract top-of-block MEV opportunities that depend on volatility and are highly sensitive to execution timing. The authors construct a dataset pairing PBS outcomes with 12-second realized volatility on centralized exchanges, demonstrating that integrated builders win blocks more frequently during periods of high volatility, confirming their advantage in capturing time-sensitive MEV.

Layer 2 rollups have experimented with alternative auction formats. Arbitrum's Timeboost mechanism implements a second-price sealed-bid auction for 'express lane' access, granting winners a time advantage in transaction sequencing \cite{timeboost2024}. Polygon's Atlas upgrade introduced a sealed-bid first-price auction format where searchers compete for per-opportunity MEV extraction rights \cite{vostrikov2025}. These sealed-bid mechanisms aim to reduce network congestion from spam-based Priority Gas Auctions (PGA) while maintaining competitive MEV extraction.

Vostrikov et al. \cite{vostrikov2025} provide an empirical analysis of atomic arbitrage on Polygon, comparing spam-based and auction-based backrunning strategies. Their findings reveal that while spam-based transactions (submitting multiple duplicate transactions at incrementally higher gas prices) remain more prevalent, auction-based strategies demonstrate superior profitability. The transition from spam to structured auctions fundamentally alters the strategic environment: searchers must now make one-shot bidding decisions under extreme time constraints without observing competitor behavior, as opposed to the iterative adjustment possible in open PGAs.

The study by Seoev et al. \cite{biddinggames2025} employs reinforcement learning to develop optimal bidding strategies for Polygon's Atlas auctions, demonstrating that history-conditioned agents can capture 49\% of available profits when deployed alongside existing searchers and 81\% when replacing the market leader. This substantially outperforms static bidding strategies derived from traditional game-theoretic equilibrium analysis, highlighting the limitations of classical auction theory in high-frequency, partially observable environments.

Zhu et al. \cite{zhu2024} examine the economic implications of revert protection, which is a feature that prevents bidders from paying fees for failed transactions. Developing a model where searchers bid for top-of-block arbitrage opportunities, they quantify how revert protection improves equilibrium auction revenue, market efficiency, and blockspace utilization. Without revert protection, optimal arbitrage strategies involve splitting opportunities into multiple small transactions to hedge against failure risk, leading to network congestion and wasted blockspace. Revert protection eliminates this inefficiency by allowing searchers to bid deterministically, knowing they will not incur costs if their transactions fail.

Empirical evidence from Ethereum Layer 2 rollups confirms these theoretical predictions of absent revert protection, with over 80\% of reverted transactions being swaps targeting high-liquidity pools (particularly USDC-WETH on Uniswap v3/v4), and searchers rarely utilize Priority Fee Auctions, instead preferring to spam duplicate transactions \cite{firstspammed2025}.

From a theoretical perspective, Mazorra et al. \cite{mazorra2022} develop a game-theoretic framework for MEV extraction, formalizing MEV games as strategic interactions where rational searchers optimize blockchain-specific constraints. They demonstrate that different ordering mechanisms, including sealed-bid auctions, first-input-first-output, and priority gas auctions, give rise to distinct equilibrium structures with varying externalities.

\section{Data}\label{sec:data}

We use the \texttt{libmev} dataset of MEV bundle transactions on Ethereum, covering the period from September 2024 to August 2025. The raw data contains 2.2 million transactions with a total extracted value of approximately \$168.5M. Each record includes the transaction hash, block number, MEV type, the tip paid to the block proposer, called a bribe, and the searcher's retained profit. We define the \textit{extracted value} of a transaction as the sum of the tip and the profit, representing the total surplus available to the searcher before any payment to the builder. Besides, we added a new feature: the \textit{bribe percentage}, the share of extracted value paid to the builder, which we use to measure competition intensity. It is computed as the ratio of the tip to the total extracted value, so that a searcher who pays most of their extraction as a bribe has a percentage near 100\%, while one who retains most of the value as private profit has a percentage near zero. Higher bribe percentages indicate fiercer competition: when many searchers compete for the same opportunity, they bid away their surplus, leaving little profit. Conversely, a low bribe percentage suggests that the winning searcher faces limited rivalry and can retain a large share of the extracted value.

\subsection{MEV Types and Market Structure}\label{subsec:mev_types_structure}

The dataset classifies transactions into four MEV types: sandwich attacks, naked arbitrage (untargeted top-of-block DEX-DEX arbitrage), backruns, and liquidations. Figure \ref{fig:mev_types} presents three views of the data disaggregated by type.

\onecolumngrid
\vspace{1em}
\begin{center}
    \includegraphics[width=0.8\textwidth]{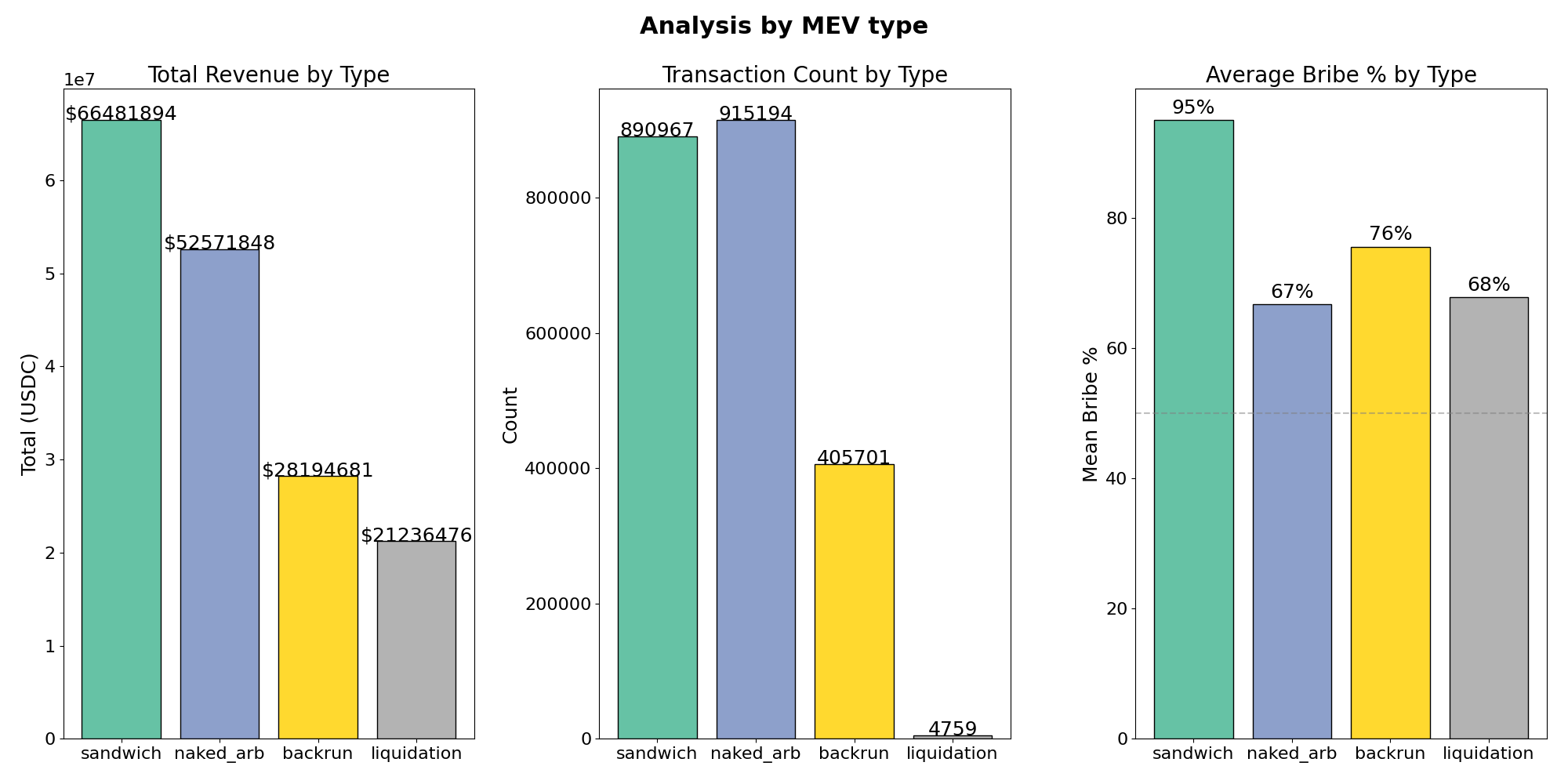}
    \captionof{figure}{\centering Analysis by MEV type. \textbf{Left:} total revenue. \textbf{Middle:} transaction count. \textbf{Right:} average bribe percentage.}
    \label{fig:mev_types}
\end{center}
\vspace{1em}
\twocolumngrid

Sandwich attacks and naked arbitrage dominate both in volume and in value. Sandwiches account for approximately \$66.5M in total extracted value across 891K transactions, while naked arbitrage generates \$52.6M across 915K transactions. Backruns contribute \$28.2M over 406K transactions, and liquidations, though they are the highest-value individual events, are comparatively rare, totaling \$21.2M across fewer than 5K transactions.

The right panel reveals substantial heterogeneity in competition intensity across MEV types. Sandwich attacks exhibit the highest average bribe percentage at 95\%, indicating that sandwich searchers operate in a fiercely competitive environment and pass nearly all of their extracted value to builders. This is consistent with the nature of sandwich attacks: the opportunity is visible in the public mempool, many searchers can detect it simultaneously, and execution is largely commoditized, leaving little room for differentiation. Naked arbitrage and liquidation searchers retain more surplus, with average bribe percentages of 67\% and 68\% respectively, suggesting that these categories require more specialized infrastructure or faster execution, which limits the effective number of competitors. Backruns fall in between at 76\%.

These cross-type differences in competition intensity have direct implications for the auction-theoretic analysis provided in Sections \ref{sec:theory} and \ref{sec:numerical}. The effective number of bidders $n$ varies across MEV types: sandwich attacks likely correspond to large $n$ (many competing searchers with similar capabilities), while liquidations and naked arbitrage correspond to small $n$ (a handful of specialized searchers). As we show in Section \ref{sec:numerical}, the linkage gap and affiliation premium are both largest for small $n$, choosing auction format most consequential precisely for those MEV categories where searchers retain the most surplus.

\subsection{Value Distribution}

Figure \ref{fig:mev_distribution} displays the distribution of extracted values across all MEV types. The left panel shows the histogram of log-transformed values, with a fitted normal density overlaid in red. The close agreement between the histogram and the fitted curve supports the log-normal specification adopted in Section \ref{sec:theory}. The right panel shows the empirical CDF on a logarithmic scale, revealing that the distribution spans over twelve orders of magnitude: from sub-cent transactions to individual events worth millions of dollars.

We fit a log-normal distribution to the pooled extracted values by maximum likelihood estimation, obtaining parameter estimates $\hat{\mu} = 1.102$ and $\hat{\sigma} = 2.524$. The large value of $\hat{\sigma}$ reflects the extreme dispersion of MEV opportunities: the standard deviation of log-values is 2.5, implying that a one-standard-deviation upward shift in the latent signal multiplies the extracted value by a factor of $e^{2.524} \approx 12.5$. This heavy-tailed structure has important consequences for the auction analysis. In particular, the order statistics of values $v_{(1)}$ and $v_{(2)}$ are highly sensitive to the number of bidders $n$: with more independent draws from a heavy-tailed distribution, the expected gap between the highest and second-highest values shrinks rapidly, which drives the strong dependence of revenue on $n$ documented in Section \ref{sec:numerical}.

The log-normal specification is a deliberate simplification. The empirical distribution exhibits slight excess kurtosis relative to the fitted normal in the left tail (visible as a minor discrepancy below $\log v \approx -5$), and the four MEV types likely have distinct distributional parameters. We pool across types for tractability and because the auction-theoretic framework in Section \ref{sec:theory} requires a single valuation distribution. A natural extension would be to estimate type-specific parameters and run the revenue simulations separately for each MEV category.

\vspace{1em}
\begin{center}
    \includegraphics[width=0.45\textwidth]{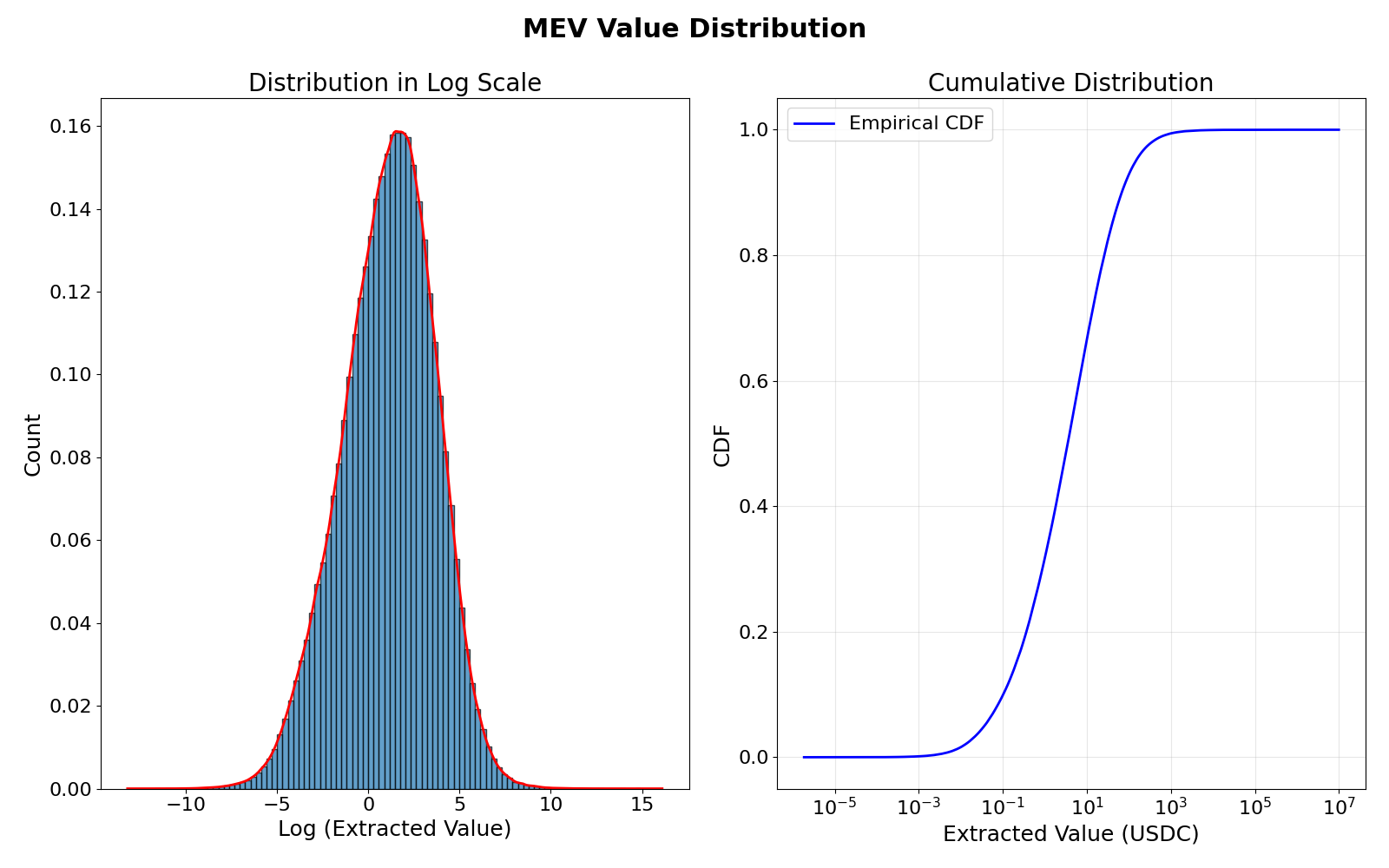}
    \captionof{figure}{\centering  MEV value distribution. \textbf{Left:} histogram of log-transformed extracted values with fitted normal density (red curve), supporting the log-normal specification. \textbf{Right:} empirical CDF on logarithmic scale, showing that the distribution spans from below $10^{-5}$ to above $10^{7}$ USDC.}
    \label{fig:mev_distribution}
\end{center}
\vspace{1em}

\subsection{Revenue Concentration}

Figure \ref{fig:concentration} documents the extreme concentration of extracted value across transactions. The left panel shows a Pareto curve: the top 1\% of transactions by value account for 68\% of total revenue, and the top 10\% account for 90\%. The right panel basically displays the same information, but as a bar chart for a better perception of statistics. The Gini coefficient of the value distribution is 0.933, confirming a level of concentration comparable to the most skewed economic distributions observed in practice.

This concentration has a practical implication for auction design: the revenue impact of mechanism choice is dominated by a small number of high-value transactions. If builders could identify high-value opportunities ex ante and apply the optimal auction format selectively, the welfare gains would be concentrated in a thin upper tail. In our simulation framework, this concentration manifests through the heavy right tail of the log-normal distribution, which generates occasional very large order statistics that drive the expected revenue differences across mechanisms.

\onecolumngrid
\vspace{1em}
\begin{center}
    \includegraphics[width=0.85\textwidth]{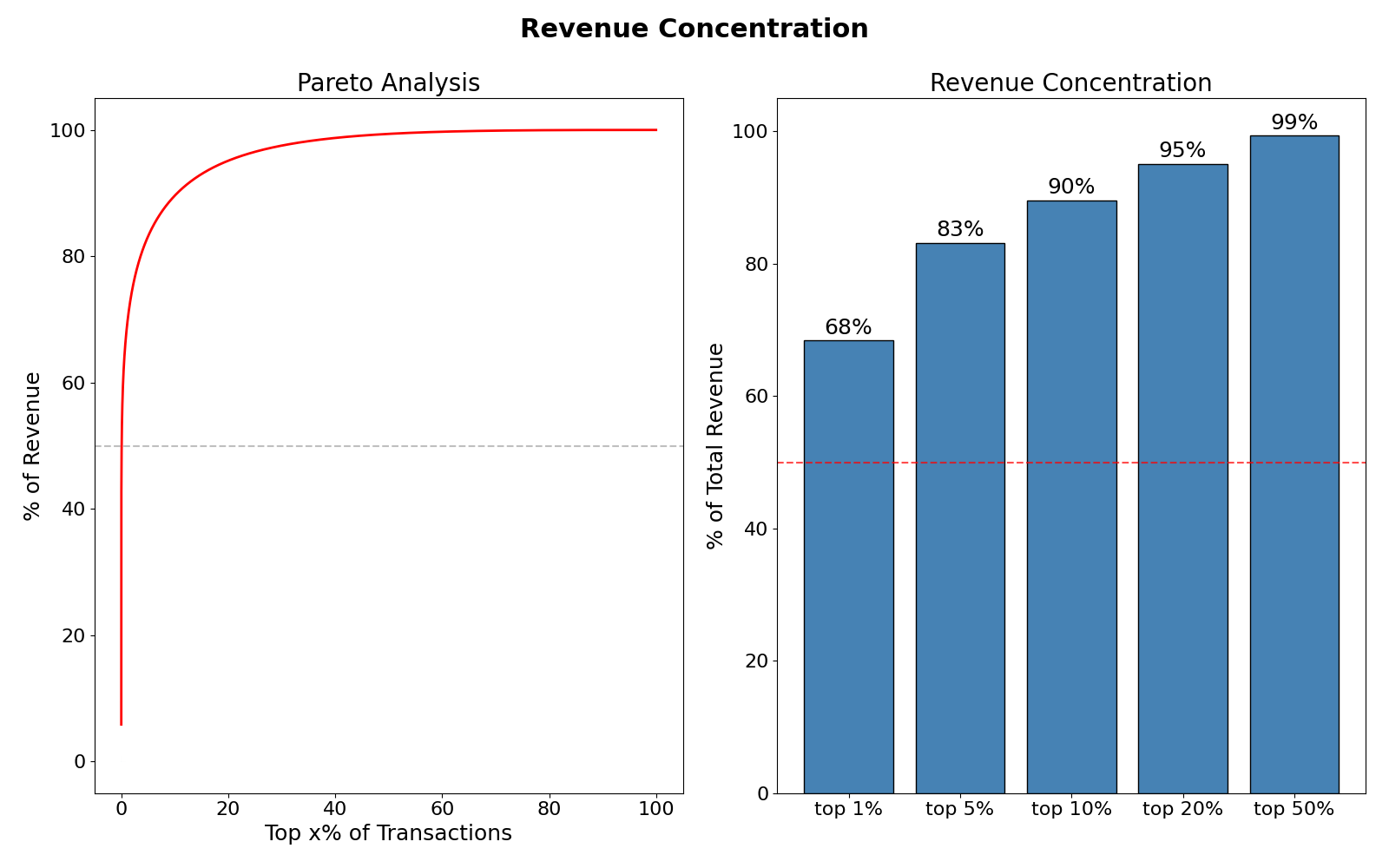}
    \captionof{figure}{\centering Revenue concentration. \textbf{Left:} Pareto curve showing cumulative revenue share as a function of the top $x$\% of transactions. The curve rises steeply: the top 1\% accounts for 68\% of total value. \textbf{Right:} revenue share of the top 1\%, 5\%, 10\%, 20\%, and 50\% of transactions. The dashed red line marks the 50\% threshold.}
    \label{fig:concentration}
\end{center}
\vspace{1em}
\twocolumngrid

\subsection{Summary Statistics}

\begin{table}[H]
\footnotesize
\begin{ruledtabular}
\begin{tabular}{ccccccc}
Type & Count & Total & Mean & Median & Std Dev & Bribe \% \\
\hline
Sandwich & 890{,}967 & 66.5 & 74.6 & 3.01 & 1{,}842 & 95\% \\
Naked arb & 915{,}194 & 52.6 & 57.4 & 3.15 & 1{,}529 & 67\% \\
Backrun & 405{,}701 & 28.2 & 69.5 & 2.28 & 2{,}104 & 76\% \\
Liquidation & 4{,}759 & 21.2 & 4{,}462 & 157.3 & 38{,}716 & 68\% \\
\hline
All & 2{,}216{,}621 & 168.5 & 76.0 & 3.01 & 1{,}925 & 79\% \\
\end{tabular}
\end{ruledtabular}
\caption{\centering Summary statistics of extracted values (USDC).}
\label{tab:table1}
\end{table}

To sum up, Table \ref{tab:table1} reports summary statistics for the pooled sample and by MEV type. Note that several features stand out. First, the mean-to-median ratio is enormous across all types: for the pooled sample, the mean (\$76.0) exceeds the median (\$3.01) by a factor of 25, which is a hallmark of heavy-tailed distributions. Second, liquidations are qualitatively different from the other types: they are rare (fewer than 5K transactions, compared to roughly 900K for sandwiches and naked arbs), but individually much larger (median \$157 vs.\ \$3 for the pooled sample). This confirms that liquidations occupy a distinct region of the $(n, \rho)$ parameter space in our auction model: low $n$, potentially high $\rho$, where the format choice matters most. Third, the standard deviation is roughly 25 times the mean for pooled data, consistent with the large $\hat{\sigma}$ estimate and the Gini coefficient of 0.933.

The bribe percentage column reinforces the competition narrative from Section \ref{subsec:mev_types_structure}: sandwiches are a near-perfectly competitive market (with 95\% bribe rate), while naked arbs and liquidations leave substantial surplus with the winning searcher, with bribe rate equal to 67--68\%. From the builders' perspective, the MEV categories with the lowest bribe percentages represent the largest potential revenue gains from improved auction design, since there is more surplus left on the table to be captured through a better mechanism.

\section{Theoretical Framework}\label{sec:theory}

Now we develop the theoretical framework for comparing different MEV auction formats. We consider a single builder who wants to sell an MEV extraction opportunity to $n$ competing searchers. Each searcher $i$ from the set of all searchers $\mathcal{N} = \{1, ..., n \}$ has a private valuation $v_i$ for the MEV opportunity. In the baseline model, valuations are drawn independently from a common distribution with cdf $F$ and pdf $f$, though we relax the independence assumption in Section \ref{subsec:affiliation}.

Our goal is to compare different auction formats in terms of equilibrium bidding strategies and their expected revenues for the builders. We compare first-price sealed bid, second-price sealed bid, all-pay, English, and Dutch auctions. For each format, we characterize the symmetric Bayesian Nash Equilibrium bidding strategy and derive the expected revenue for the builder. 

\textbf{Notation.} Throughout the paper we use $v_{(1)} \ge v_{(2)} \ge ... \ge v_{(n)}$ to denote the order statistics of bidders' valuations. Additionally, let $y_1 = \max_{j \ne i} v_j$ be the highest rival value from the perspective of bidder $i$. We write $\beta: \mathbb{R}_+ \to \mathbb{R}_+$ for the symmetric equilibrium bidding function, which maps valuations to bids.

\textbf{Distribution Specification.} Based on our empirical analysis in Section \ref{sec:data}, we adopt the log-normal distribution as the baseline valuation distribution: $v \sim \text{Lognormal} (\mu, \sigma^2)$, with density 

\[f(v) = \frac{1}{v \sigma \sqrt{2 \pi}} \exp \biggl(-\frac{(\log v - \mu)^2}{2 \sigma^2} \biggr), \quad v \in \mathbb{R}_{++}\]

\noindent and cdf equals to $F(v) = \Phi \bigl(\frac{\log v - \mu}{\sigma} \bigr)$, where $\Phi(\cdot)$ is the standard normal cdf. The lognormal specification is motivated by two features of the data: the extreme right skewness of MEV bundle values and the approximate normality of log-transformed values. Recall that the MLE estimates from Section \ref{sec:data} are $\hat{\mu} = 1.102$ and $\hat{\sigma} = 2.524$. Now we proceed with brief characterizations of five types of auctions and their equilibrium analyses, which are also shown in \cite{krishna2009}.

\subsection{First-price sealed-bid auction}
\label{subsec:fpsb}

In the first-price sealed-bid auction, each searcher submits a bid independently, the highest bid wins, and the winner pays her own bid. Searchers optimally shade their bids below their true valuations, balancing the trade-off between winning probability and surplus conditional on winning.

\textbf{Equilibrium characterization.} We derive the symmetric Bayesian Nash equilibrium in which all searchers use a common, strictly increasing bid function $\beta(v)$. Searcher $i$ with valuation $v_i$ chooses bid $b_i$ to maximize expected payoff:
\[\pi_i(b_i) = F^{n-1} (\beta^{-1} (b_i)) (v_i - b_i) \to \max_{b_i}\]
\noindent The probability of winning, $F^{n-1}(\beta^{-1}(b_i))$, follows from the fact that $\beta$ is increasing: searcher $i$ wins if and only if $b_i > \max_{j \ne i} \beta(v_j)$, which is equivalent to $\beta^{-1}(b_i) > \max_{j \ne i} v_j$, and since rival values are i.i.d.\ draws from $F$, this event has probability $F^{n-1}(\beta^{-1}(b_i))$. The first-order condition yields:
\begin{equation*}
\begin{split}
\frac{(n-1) F^{n-2}(\beta^{-1} (b_i)) f(\beta^{-1} (b_i))}{\beta' (\beta^{-1} (b_i))} (v_i - b_i) - \\ - F^{n-1} (\beta^{-1}(b_i)) = 0 
\end{split}
\end{equation*}
Imposing the symmetric equilibrium condition $b_i = \beta(v_i)$ (so that $\beta^{-1}(b_i) = v_i$) simplifies this to the ODE:
\[(n-1) F^{n-2}(v_i) f(v_i) (v_i - \beta (v_i))= F^{n-1}(v_i) \beta'(v_i) \]
With the boundary condition $\beta(0) = 0$ (the lowest type earns zero surplus), the solution is:
\begin{equation}
\begin{split}
\beta(v_i) = \frac{\int^{v_i}_0 y (n-1) F^{n-2} (y) f(y) dy}{F^{n-1}(v_i)} = \\
= v_i - \frac{\int^{v_i}_0 F^{n-1} (y) \, dy}{F^{n-1} (v_i)}
\end{split}
\label{eq:beta_fpsb}
\end{equation}

\noindent where the second equality follows from integration by parts. The equilibrium bid equals the expected value of the highest rival conditional on winning: $\beta(v_i) = \mathbb{E}[y_1 | y_1 < v_i]$. The shading $v_i - \beta(v_i) = \int_0^{v_i} F^{n-1}(y)\, dy \,/\, F^{n-1}(v_i)$ decreases as $n$ grows, since more competition reduces the gap between $v_i$ and $y_1$ conditional on winning.

\textbf{Expected revenue.} The builder's expected revenue under the first-price sealed-bid auction mechanism is $\mathbb{E}[\beta(v_{(1)})]$: the expected equilibrium bid of the winning searcher. Since $\beta(v) = \mathbb{E}[y_1 | y_1 < v]$, the law of iterated expectations gives $\mathbb{E}[\beta(v_{(1)})] = \mathbb{E}[v_{(2)}]$, establishing revenue equivalence with the second-price auction (see Section \ref{subsec:spsb}).

\subsection{Second-price sealed-bid auction}
\label{subsec:spsb}

In the second-price sealed-bid auction, each searcher submits a bid independently, the highest bid wins, but the winner pays the second-highest bid. This payment rule yields a remarkably clean equilibrium.

\textbf{Equilibrium characterization.} We claim that $\beta(v) = v$, which is the truthful bidding, is a weakly dominant strategy. To see this, consider a searcher with valuation $v$ facing a highest rival bid $\tilde{b}$. The searcher's payoff is $v - \tilde{b}$ if they win and zero otherwise.

Suppose the searcher bids $b > v$. If $\tilde{b} \in (v, b)$, the searcher wins but pays $\tilde{b} > v$, earning negative surplus, which is strictly worse than losing and earning zero. Suppose, instead, the searcher bids $b < v$. If $\tilde{b} \in (b, v)$, the searcher loses despite the existence of a profitable winning opportunity, since $v - \tilde{b} > 0$. Bidding exactly $v$ avoids both pitfalls: the searcher wins if and only if it is profitable to do so, and never overpays. This argument is independent of the distribution of rival bids or the number of competitors, making truthful bidding a weakly dominant strategy.

The outcome of the SPSB auction is the efficient allocation: the highest-value searcher wins and pays $v_{(2)}$, the second-highest valuation.

\textbf{Expected Revenue.} According to the outcome, the expected revenue of the builder equals $\mathbb{E}[v_{(2)}]$, the expected second-order statistic of $n$ draws from $F$. For the log-normal distribution, this does not admit a closed form. Hence, substituting $z = (\log v - \mu)/\sigma$ yields:
\begin{equation}
\begin{split}
\mathbb{E}[v_{(2)}] = n(n-1) \int_{-\infty}^{\infty} e^{\mu + \sigma z} \times \\
\times \ \Phi(z)^{n-2}\, \bigl(1 - \Phi(z)\bigr) \cdot \phi(z)\, dz
\end{split}
\label{eq:ev2}
\end{equation}
\noindent where $\phi(\cdot)$ is the standard normal density.

\textbf{Revenue Equivalence Theorem.} A classical result in auction theory, formalized by \cite{krishna2009}, states that under independent private values with symmetric, risk-neutral bidders, any two auction formats that (i) allocate the good to the highest-value bidder and (ii) give zero expected surplus to the lowest type yield identical expected revenue.

Both the FPSB and SPSB auctions satisfy these conditions. In SPSBA, the highest-value bidder wins and pays $v_{(2)}$. In FPSBA, the highest-value bidder wins and pays $\beta(v_{(1)}) = \mathbb{E}[y_1 | y_1 < v_{(1)}]$. Taking expectations:
\[
\mathbb{E}[\beta(v_{(1)})] = \mathbb{E}\bigl[\mathbb{E}[y_1 | y_1 < v_{(1)}]\bigr] = \mathbb{E}[v_{(2)}]
\]
since, conditional on $v_{(1)}$ being the maximum, $y_1 = v_{(2)}$. This identity extends to all standard auction formats under IPV, as we verify for the all-pay auction below.

\subsection{All-pay auction} \label{subsec:allpay}

In an all-pay auction, every searcher pays their bid regardless of whether they win. While less common in practice, this format serves as a useful theoretical benchmark and captures situations where searchers expend resources (computational effort, gas costs) that are not refunded upon losing.

\textbf{Equilibrium characterization.} Again, as in the case of FPSBA, the probability of searcher $i$ winning is the same and equals $F^{n-1}(\beta^{-1}(b_i))$. The key difference is that the bid is paid unconditionally, so the expected payoff is:
\[\pi_i(b_i) = F^{n-1} (\beta^{-1} (b_i)) v_i - b_i \to \max_{b_i}\]
The first-order condition, evaluated at the symmetric equilibrium $b_i = \beta(v_i)$, gives:
\[\beta ' (v_i) = v_i(n-1) F(v_i)^{n-2} f(v_i)  \]
Hence, we integrate this expression with the boundary condition $\beta(0) = 0$:
\begin{equation}
\beta(v_i) = \int_0^{v_i} y\, (n-1)\, F^{n-2}(y)\, f(y)\, dy
\label{eq:beta_allpay}
\end{equation}
Comparing with \eqref{eq:beta_fpsb}, the all-pay bid equals the first-price sealed-bid auction bid times the probability of winning: $\beta_{\text{AP}}(v_i) = \beta_{\text{FPSB}}(v_i) \cdot F^{n-1}(v_i)$. This relationship is intuitive: in FPSBA, the searcher pays only upon winning, so the equilibrium bid is higher per dollar, while in the All-pay auction, the searcher always pays, so the bid is scaled down by the winning probability.

\textbf{Expected revenue.} The total revenue is the sum of all bids: $R = \sum_{i=1}^{n} \beta(v_i)$. By symmetry, $\mathbb{E}[R] = n\, \mathbb{E}[\beta(v)]$. Substituting \eqref{eq:beta_allpay} and applying the identity $n(n-1) F^{n-2}(y) f(y) = \frac{d}{dy}[n F^{n-1}(y) - (n-1) F^n(y)]$, integration by parts yields $\mathbb{E}[R] = \mathbb{E}[v_{(2)}]$, confirming revenue equivalence with FPSB and SPSB auctions under IPV setting.

\subsection{Open auctions}\label{subsec:open}

We now turn to dynamic formats with different information revelation properties. Under independent private values, both open formats reduce to their sealed-bid counterparts. The distinction becomes consequential under affiliated values, as we discuss in Section \ref{subsec:linkage}.

\textbf{English auction.} In the English (or ascending) auction, the price rises continuously, and searchers drop out when the price exceeds their willingness to pay. The auction ends when the second-to-last bidder exits, and the remaining bidder wins at the dropout price.

Under IPV, the strategic analysis is straightforward: remaining active until the price reaches one's valuation $v_i$ is weakly dominant. Dropping out earlier forfeits profitable winning opportunities, while staying longer risks winning at a price above $v_i$. The auction therefore ends at price $v_{(2)}$, and expected revenue equals $\mathbb{E}[v_{(2)}]$, which is the same as a second-price sealed bid auction. This equivalence is a consequence of the dominant strategy: since $\beta(v) = v$ in both formats, the winner's identity and the payment are identical.

\textbf{Dutch auction.} In the Dutch (or descending) auction, the price starts at a high level and decreases continuously until a searcher accepts the current price. The first searcher to claim wins and pays the claiming price.

Under IPV, the Dutch auction is strategically equivalent to FPSBA. Each searcher faces the same decision problem: to choose a price at which to claim without observing rivals' actions, because the descending clock reveals no information before the claiming decision. The equilibrium bidding strategy is therefore identical to \eqref{eq:beta_fpsb}, and expected revenue equals $\mathbb{E}[v_{(2)}]$ by revenue equivalence. The strategic equivalence between Dutch and FPSB auctions holds regardless of the correlation structure, since it is driven by the absence of information revelation, not by the independence of values.

\subsection{Affiliation and the Common Factor Model}\label{subsec:affiliation}

The revenue equivalences established above rely on the assumption that searchers' valuations are independent. In MEV markets, this assumption is unlikely to hold: if market conditions (gas prices, DEX liquidity, cross-pool price discrepancies) make an arbitrage opportunity valuable to one searcher, they likely make it valuable to others. We now relax the independence assumption by introducing affiliated valuations through a Gaussian common factor model.

\textbf{Common factor structure.} Each searcher $i \in \{1, \ldots, n\}$ receives a latent signal:
\begin{equation}
z_i = \sqrt{\rho}\, Z + \sqrt{1 - \rho}\, \varepsilon_i
\label{eq:common_factor}
\end{equation}
where $Z\sim \mathcal{N}(0, 1)$ is a common factor and $\varepsilon_i \sim \mathcal{N}(0, 1)$ are i.i.d. idiosyncratic components. The valuation is $v_i = \exp(\mu + \sigma z_i)$. Since $z_j \sim \mathcal{N}(0, 1)$ marginally, the marginal distribution $v_j \sim \text{Lognormal}(\mu, \sigma^2)$ is preserved for all $\rho$, ensuring that comparisons across affiliation levels isolate the effect of correlation.

Note that the parameter $\rho \in [0, 1)$ controls the degree of affiliation: $\rho = 0$ recovers the IPV case, while $\rho \to 1$ implies near-common values. The pairwise correlation between signals is $\text{corr}(z_i, z_j) = \rho$ for $i \ne j$. In MEV markets, the common factor $Z$ captures shared information: the size of a victim transaction (sandwiches), the cross-DEX price discrepancy (arbitrage), or the collateral shortfall (liquidations), while $\varepsilon_i$ reflects differences in execution efficiency and infrastructure.

\textbf{Conditional structure.} The posterior of the common factor given searcher $i$'s signal $z_i = z^*$ is:
\begin{equation}
Z|z_i = z^* \sim \mathcal{N}\bigl(\sqrt{\rho}\, z^*,\; 1 - \rho\bigr)
\label{eq:posterior_Z}
\end{equation}
Conditional on $Z$, rival signals are independent: $z_j | Z \sim \mathcal{N}(\sqrt{\rho}\, Z,\; 1 - \rho)$ for $j \ne i$. Crucially, rivals are not conditionally independent given $z_i$ alone, as they remain correlated through the unobserved common factor $Z$. This inter-rival correlation must be accounted for in the FPSBA equilibrium computation.

\subsection{The Linkage principle}\label{subsec:linkage}

Under affiliation, the revenue equivalences established in Sections \ref{subsec:fpsb}-\ref{subsec:open} break down, and the choice of auction format becomes consequential. The key distinction is not between dynamic and static formats per se, but between formats that reveal information during the auction and those that do not.

\cite{milgrom1982} formalized this phenomenon as the \textit{linkage principle}: when values are affiliated, the seller's revenue is higher in formats that 'link' the payment to information correlated with the winner's value. The English auction achieves this linkage most effectively because bidders observe rivals' dropout prices and can update their beliefs about the intensity of competition. The SPSB auction also benefits from the linkage, as the payment $v_{(2)}$ is correlated with $v_{(1)}$ through affiliation, but to a lesser extent than the English format in general affiliated-values models. However, in our private values setting, where each searcher knows $v_i$ with certainty and learning rivals' signals does not change one's own valuation, the dominant strategy argument from Section \ref{subsec:spsb} carries through unchanged: $\beta(v) = v$ remains weakly dominant in both English and SPSB auctions regardless of affiliation. Consequently, both formats yield the same revenue, equal to the second-highest value drawn from the affiliated joint distribution.

In the FPSB and Dutch auction formats, by contrast, no information is transmitted before the bidding decision: in FPSBA, the bids are sealed, and in the Dutch auction, the descending clock reveals nothing until the first searcher claims. Despite the Dutch auction being a dynamic format, it is strategically equivalent to the static FPSBA under any correlation structure, since the absence of information revelation, not the format's timing, is what determines equilibrium behavior.

The linkage principle then implies a strict revenue ranking across three mechanism groups. The truthful-bidding formats (that is, English and SPSB auctions) generate the highest revenue, followed by the strategic-shading formats (that is, Dutch and FPSB), with the all-pay auction, computed only under IPV, at the bottom. The three levels coincide if and only if $\rho = 0$, recovering revenue equivalence under independence. The first gap, between truthful-bidding and strategic-shading formats, arises because the payment in English and SPSB auction formats is the affiliated second-order statistic, which is directly linked to the winner's value through the correlated order statistics, while the Dutch and FPSB auction formats' bids do not fully capture this linkage due to equilibrium shading. The second gap, between strategic-shading formats under affiliation and the IPV benchmark, arises because affiliation shifts the conditional distribution of rivals upward, causing bidders to shade less than they would under independence.

The magnitude of both effects depends on two factors: the degree of affiliation $\rho$ and the number of bidders $n$. Higher $\rho$ strengthens the correlation and amplifies both the linkage gap, which is the revenue gap between truthful-bidding and strategic-shading formats, and the affiliation premium, which is the gap between strategic-shading formats under affiliation and the IPV benchmark. More bidders increase the amount of information available for aggregation, but also dilute each bidder's informational contribution, so the linkage gap tends to shrink with competition while the overall revenue level rises. We quantify both effects numerically in Section \ref{sec:numerical}.

\subsection{Equilibrium under affiliation}\label{subsec:aff_equil}

To compute revenues under affiliated values, we need to understand how the common factor model from Section \ref{subsec:affiliation} changes the equilibrium behavior in each auction format. The key insight is that affiliation affects the three mechanism groups differently, because each group relies on a different component of the joint distribution.

\textbf{English and SPSB auctions.} As established in Section \ref{subsec:linkage}, truthful bidding remains weakly dominant under affiliation in the private values setting. The revenue computation is therefore straightforward: we draw $n$ affiliated valuations from the common factor model, sort them, and record the second-highest value $v_{(2)}$. Affiliation enters only through the joint distribution of order statistics, as it compresses the gap between $v_{(1)}$ and $v_{(2)}$ by pulling rival values toward the winner's value through the shared common factor $Z$. No equilibrium bid function needs to be computed for these formats.

\textbf{Dutch and FPSB auctions.} Under affiliation, the FPSBA equilibrium bid function changes because the conditional distribution of rival values, given one's own value, shifts upward. A searcher who observes a high private value $v$ infers (through the common factor) that rivals also likely have high values, which makes competition tougher than under IPV. In response, the searcher shades less: $\beta_{\text{aff}}(v) > \beta_{\text{IPV}}(v)$ for all $v$ in the interior of the support.

Following Chapter 6 of \cite{krishna2009}, the affiliated equilibrium bid function $\beta_{\text{aff}}(v)$ is characterized by a first-order ODE involving the hazard rate of the highest rival value, evaluated along the diagonal where the rival ties with the bidder. Intuitively, this hazard rate measures how likely it is that a rival is 'just below' the bidder's value: a higher hazard rate means tighter competition, which induces less shading. Under affiliation, this hazard rate increases relative to IPV because the common factor pulls rivals' values toward the bidder's own value, making near-ties more probable.

The ODE admits an integral solution that expresses the bid as the valuation minus a shading term. The shading term is a weighted average over lower values, where the weights decay exponentially in the cumulative hazard rate. In practice, evaluating the hazard rate requires computing the conditional distribution of the highest rival value given the bidder's signal, which involves integrating over the posterior of the common factor $Z$. These integrals do not admit closed-form expressions, so we evaluate them numerically by Monte Carlo integration, drawing samples from the posterior of $Z$ at each grid point. The resulting bid function is computed on a fine grid and interpolated with a cubic spline for evaluation at arbitrary values. 

\textbf{All-pay (as IPV benchmark).} The all-pay equilibrium under affiliated values has not been characterized in closed form for general affiliation structures. We therefore compute all-pay revenue under IPV only, drawing independent valuations and recording $v_{(2)}$. This provides a constant baseline that does not vary with $\rho$, against which we measure the revenue gains from affiliation in the other formats.

\textbf{Consistency check.} When $\rho = 0$, the affiliated FPSBA bid function should reduce to the standard IPV expression from Section \ref{subsec:fpsb}, and all three revenue levels should coincide by the Revenue Equivalence Theorem. We verify both properties numerically in Section \ref{sec:numerical}.

\section{Numerical Analysis}\label{sec:numerical}

We simulate expected revenues for all five auction formats, using equilibrium characterizations from Section \ref{sec:theory} and the properties of the empirical distribution from Section \ref{sec:data}. We used the simulation grid with two main hyperparameters: the number of bidders $n$ and the affiliation value $\rho$. The simulation grid spans for $n \in \{2, 3, 4, 5, 6, 7, 8, 10, 12, 15,20\}$ and $\rho \in \{0.0, 0.1, 0.2, ..., 0.9\}$ with $10^6$ auction draws per cell. The choice of grid was justified by the need to cover the full range of empirically relevant scenarios: while $n=2,..., 5$ corresponds to specialized MEV types such as liquidations or naked arbitrage, where a handful of sophisticated searchers dominate, $n=10,...,20$ would capture more competitive categories like sandwich attacks. Additionally, the affiliation parameter $\rho$ ranges from pure independence, i.e. $\rho =0$ or independent private values case, to near-common values, i.e. $\rho=0.9$, reflecting varying degrees of shared information among searchers.

For each $(n, \rho)$ pair, we compute three distinct revenue levels that correspond to groups of strategically equivalent auction formats. First of all, both English and SPSB auction revenues were estimated as the average second-highest value from affiliated draws. As it was established in Section \ref{sec:theory}, both formats share the same dominant strategy (i.e., truthful bidding) and payment rule under private values. Second, revenues of Dutch and FPSB auctions were computed applying the ODE-based equilibrium bid function from Section \ref{sec:theory} to each simulated winner. The strategic equivalence of these two formats follows from the absence of information revelation in the Dutch format before the claiming decision. Third, all-pay auction revenue was estimated as the IPV benchmark from independent draws. Since the all-pay equilibrium under affiliation has not been characterized in closed form for general affiliation structures, we compute it under IPV only, providing a constant baseline that does not vary with $\rho$. 


\subsection{Revenue Equivalence and Linkage Principle Verification}

We verify the simulation against two fundamental theoretical benchmarks before reporting the main results. These checks serve a dual purpose: they confirm that the numerical implementation is correct and establish the precision of the Monte Carlo estimates.

Under IPV, or in our setting if $\rho=0$, the Revenue Equivalence Theorem requires all standard auction formats to yield the same expected revenue, provided the symmetric equilibrium assigns the object to the highest-value bidder and gives zero surplus to the lowest type. To check the correctness of computations for all types of auctions, we have made a simulation of all given formats for the case when $\rho =0$. Table \ref{tab:table2} confirms the agreement within 0.2\% across all tested values of $n$. The residual differences are consistent with Monte Carlo sampling noise at $N = 10^6$ draws and would vanish in expectation as $N \to \infty$.

\begin{table}[H]
\begin{ruledtabular}
\begin{tabular}{ccccc}
$n$ & English / SPSBA & Dutch / FPSBA & All-pay & 
$\left|\dfrac{\text{FPSBA} - \text{Eng}}{\text{Eng}}\right|$ \\
\hline
2  & 5.41  & 5.41  & 5.42  & 0.1\% \\
5  & 30.02 & 30.06 & 29.85 & 0.1\% \\
10 & 79.72 & 79.89 & 79.57 & 0.2\% \\
\end{tabular}
\caption{Verification of revenue equivalence at $\rho = 0$.}
\label{tab:table2}
\end{ruledtabular}
\end{table}

One can note that the absolute revenue levels grow rapidly with $n$: from \$5.41 at $n = 2$ to \$79.72 at $n=10$. This reflects the heavy tail of the empirical MEV distribution, described in Section \ref{sec:data}, which implies that $v_{(2)}$ (e.g., second-highest valuation) is highly sensitive to the number of competing draws. With 10 independent log-normal draws, the second-order statistic already reaches the far right tail of the distribution, generating large expected payments. Additionally, one can note that the revenue discrepancies between the three auction formats grow with $n$, though we can treat them as negligible.  

Speaking of the linkage principle, let us take a look at the case when $\rho = 0.5$. Under affiliation, \cite{milgrom1982} established that the linkage principle predicts the strict ranking where the English auction gains the same revenue as SPSBA but higher revenue than Dutch auction and FPSBA, as they are equivalent in the sense of revenue even under affiliation. The least amount of revenue in this hierarchy gains the all-pay auction. Our test verifies the more delicate component of the simulation: the ODE-based computation of $\beta_{\text{aff}}$, which requires accurate evaluation of the conditional hazard rate along the diagonal. To sum up, Table \ref{tab:table3} confirms the strict ranking, with English/SPSBA exceeding Dutch/FPSBA by 14-28\%, while Dutch/FPSBA exceeding the IPV benchmark (All-pay auction) by 41-122\%.

First of all, it is worth paying attention to the following results: the gap between English and FPSB auctions decreases with $n$ (from 27.7\% in the case when $n=2$ to 14.0\% when $n=10$), which is consistent with the theoretical intuition that the linkage advantage dilutes with competition. When many bidders compete, each individual's information is less pivotal, and the sealed-bid format's revenue loss from not aggregating information diminishes. Furthermore, the gap between FPSB and All-pay auctions also decreases with $n$ (from 122.2\% in the case when $n = 2$ to 40.7\% when $n = 10$), reflecting the fact that affiliation-induced bid adjustment in the FPSBA format is proportionally larger when fewer bidders compete, and each bidder conditions more heavily on their own signal about the common factor. These monotonicity properties hold across the full grid, as we show below.

\onecolumngrid
\vspace{1em}
\begin{center}
\begin{tabular}{cccccc}
\hline\hline
$n$ & English / SPSBA & Dutch / FPSBA & All-pay (IPV) & Linkage gap & FPSBA vs AP \\
\hline
2  & 15.24  & 11.93  & 5.37  & 27.7\% & 122.2\% \\
5  & 59.87 & 51.10 & 30.04 & 17.2\% & 70.1\% \\
10 & 128.21 & 112.48 & 79.93 & 14.0\% & 40.7\% \\
\hline\hline
\end{tabular}
\captionof{table}{\centering Verification of linkage principle at $\rho = 0.5$. Linkage gap $= (\text{Eng} - \text{FPSBA})/\text{FPSBA}$; last column $= (\text{FPSBA} - \text{AP})/\text{AP}$.}
\label{tab:table3}
\end{center}
\vspace{1em}
\twocolumngrid

What follows is a mechanism comparison in terms of the other hyperparameter: the number of bidders, $n$. Table \ref{tab:table4} compares all five formats at $n = 5$, which is a plausible bidder count for specialized MEV types, such as liquidations and naked arbitrage. One can note that at $\rho=0$, revenue equivalence holds: all formats yield approximately \$30. As affiliation increases, the three revenue levels fan out. At $\rho = 0.5$, both English auction and SPSBA generate \$59.87, which is 99\% above the IPV benchmark, while Dutch auction and FPSBA generate \$51.10, which is 70\% above the IPV benchmark, and the All-pay baseline remains flat at approximately \$30. The English/SPSBA advantage over Dutch/FPSBA is 17\% at $\rho=0.5$ and grows to 18\% at $\rho=0.8$.

Table \ref{tab:table4} reveals three distinct auction groups. The first group, English auction and SPSBA, produces identical revenues across all columns, confirming that the dominant strategy argument ($\beta (v) = v$) holds regardless of correlation. Both formats generate the second-highest draw from the affiliated distribution, and since our model is private values, the additional price discovery in the English format provides no informational advantage to bidders over SPSBA. The second group, FPSB and Dutch auctions, also produces identical revenues, confirming the strategic equivalence of sealed-bid and descending formats. Their revenues are systematically below English/SPSBA for $\rho > 0$, reflecting the bid shading inherent in the first-price equilibrium: even though affiliation causes bidders to shade less than under IPV, some shading persists, and the resulting revenue loss is the linkage gap. The third group, All-pay auction as the IPV benchmark, remains constant at \$30 across all $\rho$ columns. This pattern is by construction as the all-pay draws are independent, but it also serves as a useful diagnostic: any systematic trend in this row would indicate an implementation error or the simulation loop.

\begin{table}[H]
\begin{ruledtabular}
\begin{tabular}{cccccc}
Auction format & $\rho = 0$ & $\rho = 0.2$ & $\rho = 0.4$ & $\rho = 0.5$ & $\rho = 0.8$ \\
\hline
English  & 30.02  & 40.05  & 53.49  & 59.87 & 83.15 \\
SPSBA  & 30.02  & 40.05  & 53.49  & 59.87 & 83.15 \\
FPSBA & 30.06 & 37.02 & 46.10 & 51.10 & 70.75 \\
Dutch & 30.06 & 37.02 & 46.10 & 51.10 & 70.75 \\
All-pay & 29.85 & 30.33 & 30.05 & 30.04 & 29.96 \\
\end{tabular}
\caption{\centering Expected revenue per auction for $n=5$ bidders across affiliation levels.}
\label{tab:table4}
\end{ruledtabular}
\end{table}

\onecolumngrid
\vspace{1em}
\begin{center}
    \includegraphics[width=\textwidth]{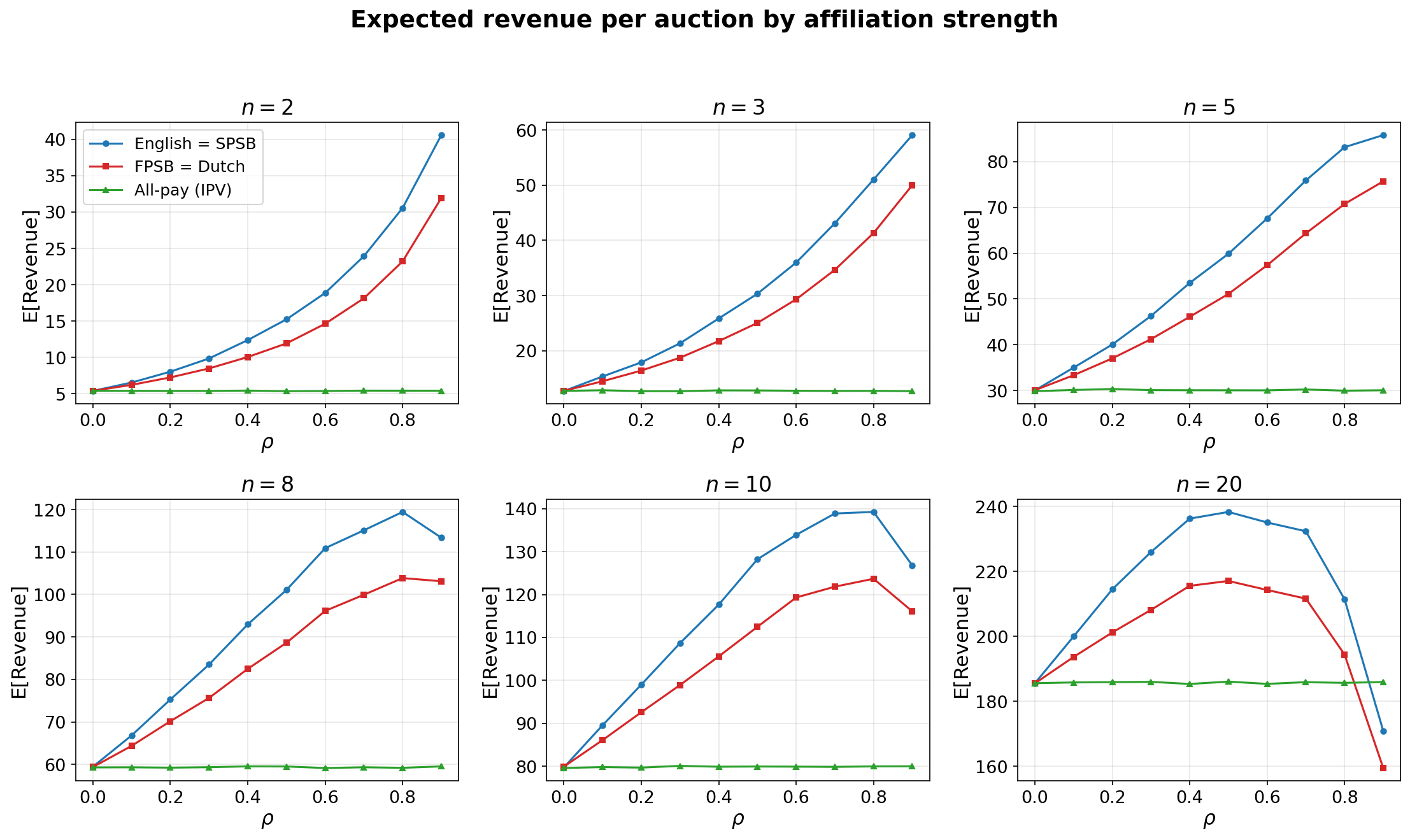}
    \captionof{figure}{\centering Expected revenue vs $\rho$ for different values of competition level, $n$. Blue line stands for English and SPSB auctions, red line stands for Dutch and FPSB auctions, and green line stands for All-pay auction (IPV benchmark, does not change with $\rho$).}
    \label{fig:revenue_vs_rho}
\end{center}
\vspace{1em}
\twocolumngrid

It is important to look into the economic magnitude of the differences. Moving from $\rho = 0$ to $\rho = 0.5$ nearly doubles revenue for both English/SPSBA and Dutch/FPSBA. This effect arises because the common factor $Z$ in the Gaussian copula model compresses the gap between $v_{(1)}$ and $v_{(2)}$: when searchers' values are positively correlated, the second-highest bidder's value is drawn closer to the winner's value, increasing the payment. At $\rho = 0.8$, English/SPSBA reaches \$83.15, which is nearly three times the IPV baseline, indicating that the affiliation premium dominates the competitive effect for this bidder count.

\subsection{Revenue Comparison}

We now examine how revenues vary across the full $(n, \rho)$ parameter space. The interaction between competition level and affiliation value produces some non-linear patterns that have direct implications for MEV mechanism design.

\textbf{Revenue and affiliation.} Figure \ref{fig:revenue_vs_rho} plots expected revenue against $\rho$ for six bidder counts. It is important to note two major patterns: firstly, for $n \le 8$, revenue increases monotonically in $\rho$ both English/SPSBA and Dutch/FPSBA pairs, with a widening gap between them. The mechanism here is intuitive: when searchers' values are positively correlated through the common factor $Z$, the second-highest value $v_{(2)}$ is drawn closer to the highest value $v_{(1)}$, so the payment, whether $v_{(2)}$ directly (as in case of English and SPSB auctions), or $\beta_{\text{aff}}(v_{(1)})$ (as in case of Dutch and FPSB auctions), increases. For $n=2$, English and SPSB auctions' revenue grows from \$5.41 at $\rho=0$ to \$40.59 at $\rho = 0.9$: a 650\% increase driven entirely by the compression of the $v_{(1)} - v_{(2)}$ gap.

Secondly, for $n = 10$, revenue peaks around $\rho= 0.6-0.7$ and then declines. At $n = 20$, English and SPSB auctions peak at \$238 at $\rho = 0.5$, and falls to \$171 at $\rho = 0.9$, even below the IPV benchmark of \$186. This reversal occurs because near-perfect correlation collapses the spread between order statistics: when all bidders observe approximately the same value, the competitive pressure that drives $v_{(2)}$ close to $v_{(1)}$ disappears. More precisely, $v_{(2)}$ under affiliation can be decomposed $\exp (\mu + \sigma(\sqrt{\rho}Z + \sqrt{1-\rho}\varepsilon_{(2)}))$, where $\varepsilon_{(2)}$ is the second-largest idiosyncratic shock. Note that as $\rho \to 1$, the idiosyncratic variance $\sigma^2 (1-\rho) \to 0$ and the order-statistic spread, which generates high $v_{(2)}$ when $n$ is large under IPV, collapses. The net effect on $\mathbb{E}[v_{(2)}]$ is non-monotone: initially, correlation helps by compressing $v_{(1)} - v_{(2)}$, but eventually it hurts by eliminating the right-tail amplification from multiple independent draws.

\textbf{Revenue and competition.} Figure \ref{fig:revenue_vs_n} plots revenue against $n$ for six different affiliation levels. At $\rho = 0$, which is depicted on the top-left panel, all three lines coincide, providing a visual confirmation of revenue equivalence. As $\rho$ increases, the three lines separate progressively: the English and SPSBA line rise above the Dutch and FPSBA line, which rises above the All-pay auction line. Revenue increases with $n$ for all formats, since more bidders push the order statistics higher. However, the rate of increase depends on $\rho$: under the IPV setting, the marginal revenue from an additional bidder is driven by the increase in $\mathbb{E}[v_{(2)}]$ from one more independent draw, while under affiliation, the additional bidder's contribution is partially offset by the correlation that makes their value redundant with existing bidders' values.

At $\rho = 0.9$, the highest level of affiliation, which is depicted on the bottom-right panel, a dramatic crossover occurs: the IPV benchmark overtakes the affiliated formats for $n\ge 15$, confirming that the revenue reversal at extreme affiliation is not an artifact of a particular $n$ but a systematic phenomenon that applies across all auction mechanisms. This has practical implications for MEV markets: in highly competitive categories like sandwich attacks, where many searchers observe similar opportunities, the correlation structure may actually suppress revenue relative to what a naive IPV model would predict.

\onecolumngrid
\vspace{1em}
\begin{center}
    \includegraphics[width=\textwidth]{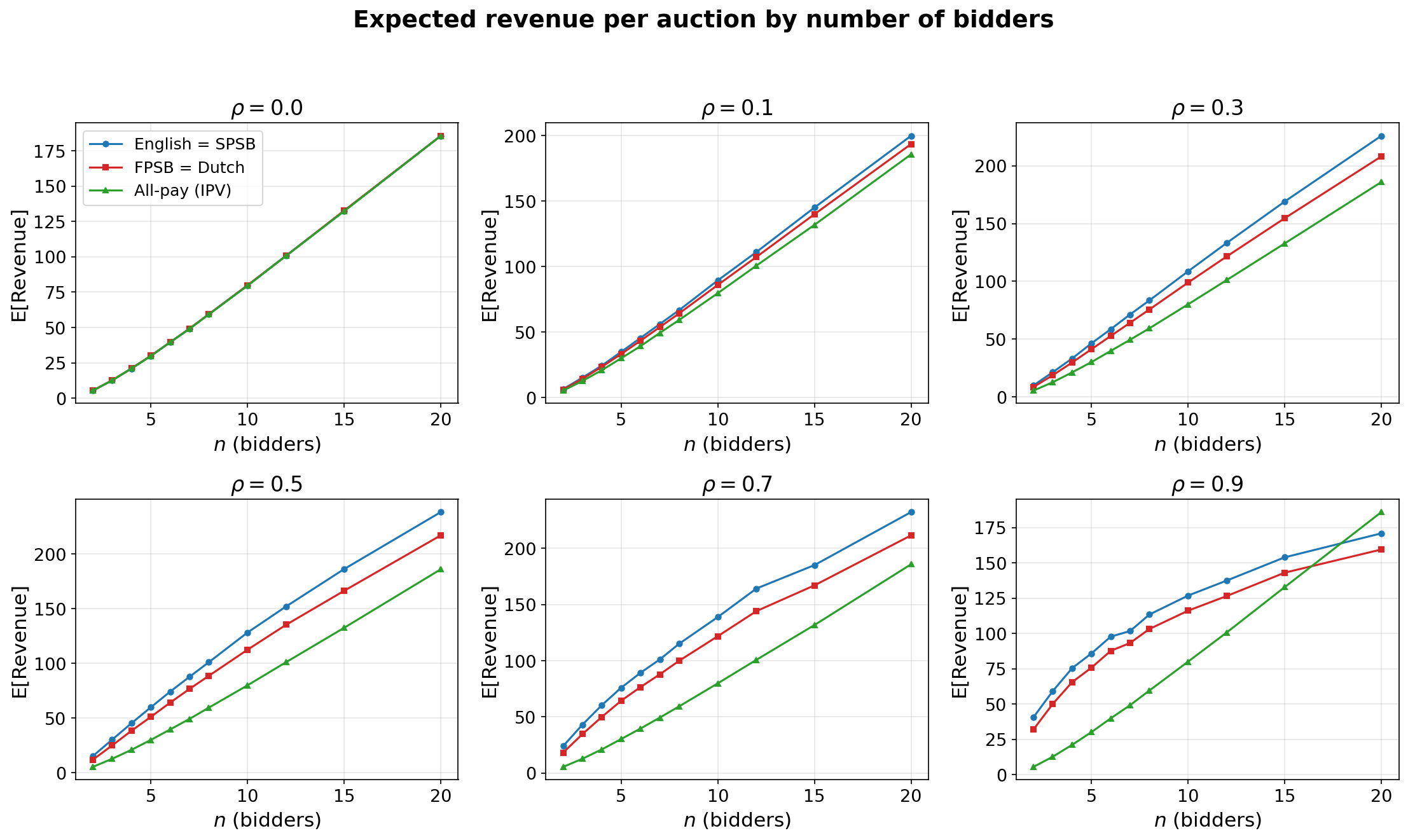}
    \captionof{figure}{\centering Expected revenue vs $n$ for different values of affiliation, $\rho$. Blue line stands for English and SPSB auctions, red line stands for Dutch and FPSB auctions, and green line stands for All-pay auction (IPV benchmark, does not change with $\rho$).}
    \label{fig:revenue_vs_n}
\end{center}
\vspace{1em}
\twocolumngrid

Tables \ref{tab:appendix_1}, \ref{tab:appendix_2}, and \ref{tab:appendix_3} in the Appendix report the complete revenue grids for each mechanism group. Several patterns visible in the figures can be confirmed numerically. In Table \ref{tab:appendix_1}, which depicts the results for English and SPSB auctions, each row increases monotonically from left to right for $n \le 8$, confirming the monotone affiliation effect for small bidder counts. For $n \ge 10$, the rows peak in the interior: $n = 10$ peaks at $\rho = 0.8$ (at \$139.24), $n = 15$ peaks at $\rho = 0.6$ (at \$192.45), and $n = 20$ peaks at $\rho = 0.5$ (at \$238.33). The peak shifts leftward as $n$ increases, indicating that the revenue-maximizing level of affiliation decreases with competition.

Table \ref{tab:appendix_2}, which shows the results for Dutch and FPSB auctions, exhibits the same qualitative patterns but at uniformly lower levels for $\rho > 0$. Comparing the two tables cell by cell, the English and SPSBA premium ranges from near-zero at $\rho = 0$ to over 30\% at small $n$ and high $\rho$.

Table \ref{tab:appendix_3}, which is the IPV benchmark, or the All-pay auction case, serves as a diagnostic: its row-wise standard deviation across $\rho$ columns is approximately 0.02--0.3, confirming negligible Monte Carlo noise. For instance, the $n = 5$ row fluctuates between 29.85 and 30.33 across the ten $\rho$ columns, a range of 1.6\% around the mean of 30.07. This noise level sets the floor for meaningful comparisons: any revenue difference smaller than approximately 0.5\% should not be interpreted as economically and statistically significant.

\subsection{Linkage gap}

The main prediction of the Linkage principle is that the open auction format generates higher revenue than the sealed-bid format under affiliated values. In order to check that we quantify this effect through the \textit{linkage gap}, which is defined as the percentage by which English or SPSBA expected revenue exceeds Dutch or FPSBA expected revenue, expressed relative to the Dutch/FPSBA level. In other words, it measures how much additional revenue (in percent) an auctioneer would gain by switching from a sealed-bid format to an open ascending format.

This metric is directly actionable for mechanism designers: it measures the revenue a builder leaves on the table by choosing a sealed-bid format over an ascending one, holding all other design parameters fixed. Figure \ref{fig:linkage_gap_heatmap} displays the gap across all $(n, \rho)$ pairs.

\onecolumngrid
\vspace{1em}
\begin{center}
    \includegraphics[width=\textwidth]{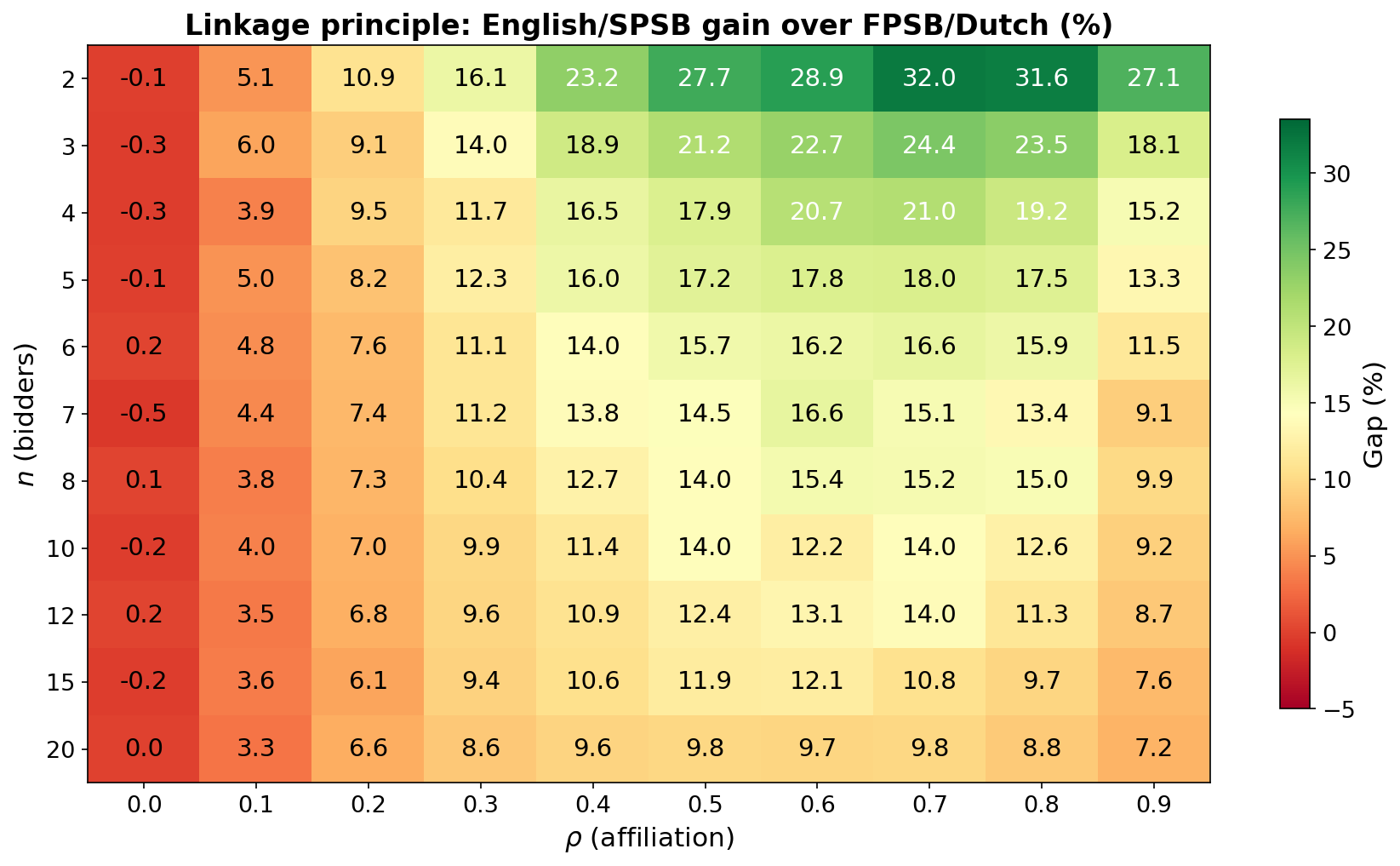}
    \captionof{figure}{\centering Linkage gap in \% across $(n, \rho)$ pairs. The color scale ranges from dark-red (for near-zero gap at $\rho = 0$) through orange and yellow (3-15\% for moderate parameters) to dark-green highlights at the peak (32\% at $(n, \rho) = (2,0.7)$. The gap is non-negative for virtually all cells, with the only exceptions being entries at $\rho = 0$ that fluctuate within $\pm 0.5 \%$ of zero due to Monte Carlo noise.}
    \label{fig:linkage_gap_heatmap}
\end{center}
\vspace{1em}
\twocolumngrid

Four regularities emerge from the heatmap. Together, they paint a consistent picture: the linkage principle holds without exception across the parameter space, the gap is economically meaningful for realistic MEV market calibrations, and its variation across $n$ and $\rho$ aligns with the theoretical intuition that the open format's advantage stems from information aggregation. We now discuss each regularity in turn.

\textit{Regularity 1: Non-negativity.} The gap is non-negative across all 110 cells in the grid. At $\rho = 0$, the entries fluctuate within $\pm 0.5\%$ of zero, consistent with Monte Carlo noise and the Revenue Equivalence prediction. For any $\rho > 0$, the gap is strictly positive. No cell violates the theoretical bound by more than the Monte Carlo noise floor, providing a clean confirmation of the Linkage principle theorem across the entire parameter space.

\textit{Regularity 2: Hump-shape in $\rho$.} For fixed $n$, the gap first increases with $\rho$, peaks around $\rho = 0.7\text{--}0.8$ for small $n$ and $\rho = 0.5\text{--}0.6$ for large $n$, and then declines at $\rho = 0.9$. The decline at extreme affiliation reflects the convergence of all valuations to a common value: when $\rho \approx 1$, both $v_{(2)}^{\text{aff}}$ and $\beta_{\text{aff}}(v_{(1)}^{\text{aff}})$ are driven by the same common factor $Z$, and the informational advantage of the open format vanishes.

\textit{Regularity 3: Monotone decrease in $n$.} For fixed $\rho > 0$, the gap decreases with $n$: at $\rho = 0.5$, it declines from 27.7\% (for $n = 2$) to 17.2\% (for $n = 5$), 14.0\% (for $n = 10$), and 9.8\% (for $n = 20$). With more bidders, competition already drives $v_{(2)}$ close to $v_{(1)}$, so the additional revenue from an open format is proportionally smaller. Equivalently, each bidder's information becomes less pivotal in a crowded auction, reducing the value of the linkage mechanism.

\textit{Regularity 4: Stable range for realistic parameters.} For the parameter range most relevant to MEV markets ($n \in [5, 10]$, $\rho \in [0.3, 0.7]$), the gap stabilizes at 10-18\%. This stability is practically important: it means that the builder does not need to precisely estimate $n$ or $\rho$ to conclude that the format choice matters. Across a wide range of plausible calibrations, the English or SPSB auction formats generate roughly 10-18\% more revenue than the Dutch/FPSBA formats.

\textbf{Dollar interpretation.} To translate the percentage gaps into dollar figures, we apply them to the observed bribe totals from Section \ref{sec:data}. With $n = 5$ and $\rho = 0.5$, the 17.2\% gap applied to observed bribes of \$101.3M implies approximately \$17.4M in foregone revenue from using sealed-bid rather than ascending auctions. Even at the conservative end ($n = 10$, $\rho = 0.3$), the 9.9\% gap corresponds to \$10.0M. These estimates should be interpreted as upper bounds for two reasons. First, they assume symmetric bidders: in practice, heterogeneous capabilities may reduce the effective competition below what $n$ suggests. Second, they assume that the builder can costlessly implement the ascending format, ignoring the latency and communication overhead that may favor the simplicity of sealed-bid mechanisms. Nonetheless, a gap of \$10-18M over the sample period represents a substantial incentive for builders to consider format redesign.

\subsection{Gain over the IPV Benchmark}

\begin{figure}[H]
    \includegraphics[width=\textwidth]{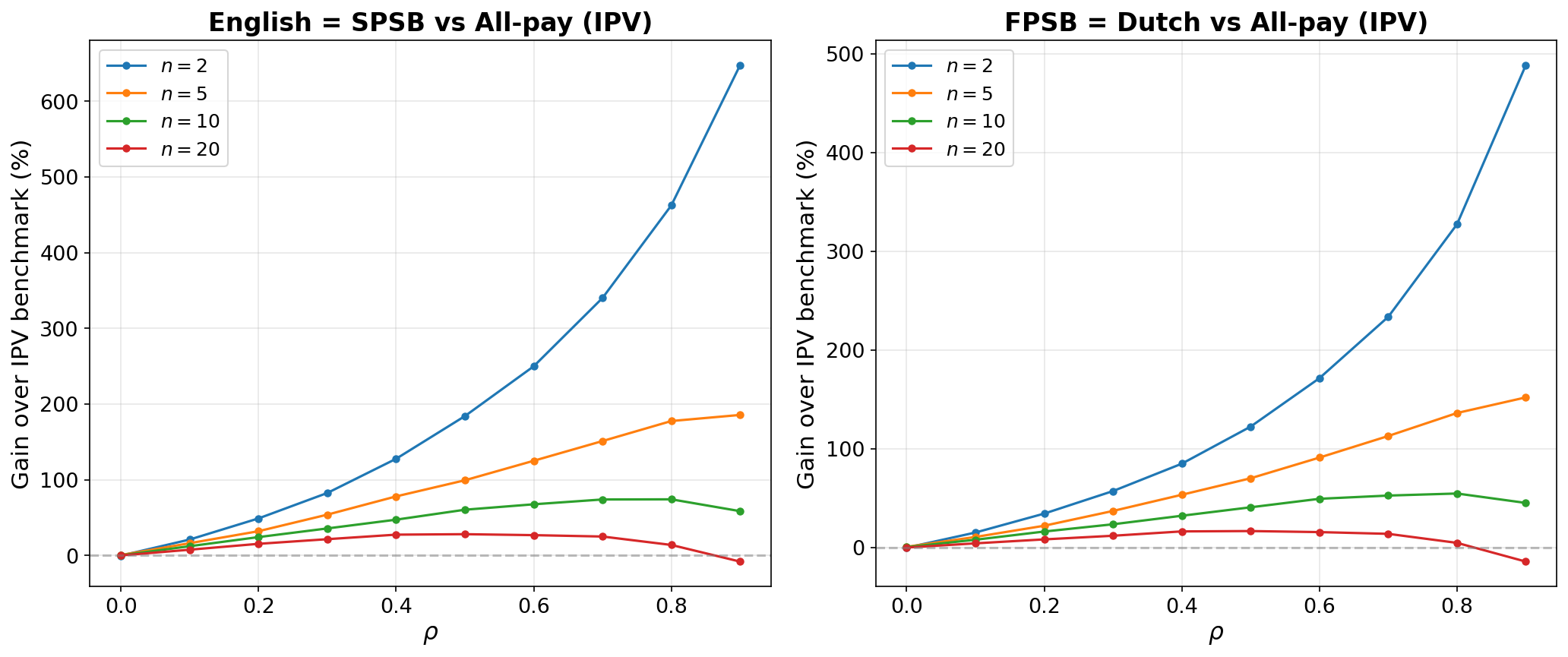}
    \begin{minipage}{\textwidth}
    \caption{\centering Revenue gain over IPV benchmark (\%). \textbf{Left:} English and SPSB auctions result. \textbf{Right:} Dutch and FPSB auctions result. Gains increase with $\rho$ and decrease with $n$, reversing at extreme $\rho$ for large $n$.}
    \label{fig:gain_vs_ipv}
    \end{minipage}
\end{figure}

The linkage gap measures the relative advantage of English/SPSBA over Dutch/FPSBA, but both affiliated formats generate substantially more revenue than the IPV benchmark. We now quantify this absolute affiliation premium: how much additional revenue does the builder capture by correctly accounting for the correlation structure of MEV valuations, relative to a model that assumes independence? Figure \ref{fig:gain_vs_ipv} plots the revenue gain of English/SPSBA and Dutch/FPSBA over the all-pay (IPV) benchmark across the $\rho$ dimension, for four selected values of $n$.

\onecolumngrid
\twocolumngrid

At $\rho = 0.5$, English and SPSB auction formats generate 28-184\% more revenue than the IPV benchmark (depending on $n$), and Dutch/FPSB auction formats generate 17-122\% more. These are first-order effects, not marginal corrections: affiliation is the primary determinant of auction revenue in the MEV context. To put these numbers in perspective, at $n = 5$ and $\rho = 0.5$, the English/SPSBA format collects \$59.87 per auction compared to \$30.04 under IPV, a near-doubling of revenue that arises purely from the correlation structure of searchers' valuations. This implies that any auction-theoretic analysis of MEV markets that assumes independent private values will systematically underestimate the revenue potential of well-designed mechanisms.

The gains are largest for small $n$, e.g., $n \approx 2-5$, precisely the regime relevant for liquidations and naked arbitrage where few sophisticated searchers compete. For $n = 2$ and $\rho = 0.5$, the English/SPSB auction format's gain is 184\%, meaning that correlation nearly triples the expected payment. The economic logic is straightforward: with only two bidders, the second-highest value $v_{(2)}$ under IPV is often far below $v_{(1)}$, especially with the heavy-tailed log-normal distribution. Affiliation pulls $v_{(2)}$ toward $v_{(1)}$ through the common factor, closing this gap and dramatically increasing the payment. As $n$ grows, the gap between $v_{(1)}$ and $v_{(2)}$ is already small under IPV (due to the compression from many independent draws), so the additional compression from affiliation has a smaller proportional impact.

At $n = 20$ and $\rho = 0.9$, both English and Dutch underperform the IPV benchmark. This reversal, which is visible as the $n = 20$ line dipping below the dashed zero line in both panels of Figure \ref{fig:gain_vs_ipv}, reflects the collapse of idiosyncratic variation at extreme affiliation. With $\rho \approx 1$, all bidders share approximately the same value $\exp(\mu + \sigma Z)$, and the order-statistic spread that generates high $v_{(2)}$ from 20 independent draws vanishes. The expectation $\mathbb{E}[\exp(\mu + \sigma Z)] = \exp(\mu + \sigma^2/2)$ is finite but does not benefit from the order-statistic amplification that arises under IPV. In practical terms, this regime, when there are many bidders with near-identical valuations, resembles a common-value auction with negligible private information, where the seller extracts the common value regardless of the mechanism.

The asymmetry between the two panels is also informative: the Dutch/FPSB auction formats gain, depicted on the right panel, is uniformly below the English/SPSB auction formats gain, depicted on the left panel, for $\rho > 0$, with the vertical distance between the two representing the linkage gap discussed in the previous subsection. This visual separation is widest at moderate $\rho$ and small $n$, and narrows at both extremes, consistent with the hump-shaped linkage gap documented in the Figure \ref{fig:linkage_gap_heatmap}.

\section{Conclusion}

This paper examines five auction formats in the context of MEV markets on Ethereum: first-price sealed-bid, second-price sealed-bid, English, Dutch, and all-pay. Using 2.2 million transactions documenting \$168.5 million in total MEV extraction across three major auctioneers (Blink, Merkle, and MEV Blocker), we establish three empirical regularities: extreme right-tail concentration (top 1\% of transactions generate 68\% of revenue), log-normal valuations with very heavy tails, and significant heterogeneity in competition intensity across MEV types. We then derive equilibrium bidding strategies and expected revenues for all five formats under both independent private values and affiliated values, modelling correlation through a Gaussian common factor. Our numerical simulations across the full $(n, \rho)$ parameter grid confirm the linkage principle: English and SPSB auctions strictly dominate Dutch and FPSB formats for any $\rho > 0$, with a linkage gap of 14-28\% at moderate affiliation and up to 30\% for small bidder counts, while the all-pay format consistently underperforms all standard formats. Applied to the observed bribe totals, these gaps correspond to \$10-18 million in foregone revenue over the sample period. We also document a novel non-monotonicity: at large $n$ and high $\rho$, revenue peaks in the interior of the parameter space and then declines.

\subsection{Practical Recipe for Auction Hosts}

The theoretical and empirical results suggest a concrete approach for builders seeking to maximize revenue from their auctions. We state it here as a practitioner's guide that synthesizes both the formal results and practical constraints that fall outside the mathematical model.

\textbf{Collect and characterize transaction data.} Before choosing a mechanism, an auctioneer should build a dataset of historical extractions with at minimum three fields: extracted value $v$, bribe payment, and MEV type. The ratio of bribe to extracted value is the key diagnostic. An auctioneer with only a few weeks of data can already estimate the tail index of the value distribution, which largely determines how much of the total revenue is concentrated in high-value lots and therefore how much the mechanism choice matters.

\textbf{Estimate the valuation distribution.} Fit a log-normal distribution to the extracted values, as motivated by the approximate normality of log-transformed values documented in Section \ref{sec:data}. The MLE estimates $(\hat{\mu}, \hat{\sigma})$ are sufficient inputs for the numerical revenue computations in Section \ref{sec:numerical}. If $\hat{\sigma} > 2$, the distribution is in the very heavy tails regime, and mechanism optimization for the right tail dominates overall revenue performance. Auctioneers operating in this regime should weigh high-value transaction outcomes more heavily in any subsequent analysis.

\textbf{Estimate affiliation $\rho$ and effective bidder count $n$.} Affiliation can be estimated from the cross-sectional dispersion of bribe percentages within a MEV type: if searchers who compete for the same opportunity tend to bid at similar levels, this is evidence of a strong common factor. Formally, the intra-class correlation of log-bids within an opportunity identifies $\rho$. The effective bidder count $n$ can be approximated from the empirical bribe distribution. Under our log-normal specification, the equilibrium shading depends on $n$, $\mu$ and $\sigma$ through the conditional order-statistic distribution, so there is no clean closed-form inversion. In practice, one can calibrate $\hat{n}$ by matching the observed mean bribe percentage to the simulated FPSB equilibrium bribe ratio from Section \ref{sec:theory}: for each candidate $n$, the simulation produces a predicted mean bribe ratio, and $\hat{n}$ is the value that minimizes the distance to the empirical bribe percentage. Practitioners should compute separate ($\hat{\rho}$, $\hat{n}$) estimates for each MEV type, as these parameters differ substantially across sandwiches, arbitrage, backruns, and liquidations.

\textbf{Choose the auction format.} Given the estimates from valuation distribution, affiliation, and effective bidder count estimates, the following decision rules follow from the analysis in Sections \ref{sec:theory} and \ref{sec:numerical}, augmented by practical considerations beyond the theoretical model.

\begin{enumerate}
    \item \textbf{Use English or SPSB if $\rho > 0.2$ and trust in auctioneer is high.} This is the main recommendation. For any positive affiliation, the Linkage principle guarantees that English and SPSB formats strictly dominate FPSB and Dutch. The gain is 14--28\% at $\rho = 0.5$ and grows further at higher correlation. SPSBA is preferable over English when latency is a binding constraint: it requires only a single sealed bid rather than a live ascending clock, reducing infrastructure demands. However, SPSBA is vulnerable to shill bidding (the auctioneer can insert a fake bid just below the winner's bid to inflate the payment), so it should only be deployed when the mechanism is implemented transparently, for instance via a verifiable on-chain contract. English auctions mitigate this risk through public price revelation but introduce collusion vulnerabilities, as bidders can coordinate through observed dropout prices.

    \item \textbf{Use FPSBA or Dutch if $\rho \approx 0$, collusion risk is high, or latency is critical.} Under near-independence, revenue equivalence holds, and the format choice does not materially affect expected payments. FPSBA eliminates real-time strategic interaction, reducing the risk of tacit collusion among repeat players, a realistic concern in MEV markets where the same small set of sophisticated searchers compete repeatedly. Dutch auctions are strategically equivalent to FPSBA and offer speed advantages (the auction terminates as soon as any bidder accepts), making them well-suited to time-sensitive MEV opportunities. Neither format is vulnerable to shill bidding in the same way as SPSB, since winners pay their own bids. However, both formats require bidders to solve a harder strategic problem (optimal bid shading) and may disadvantage less sophisticated searchers, reducing effective competition.

    \item \textbf{Avoid all-pay formats in MEV contexts.} Despite the theoretical revenue premium of all-pay auctions under symmetric IPV, our simulations show that FPSB revenue exceeds all-pay by 40--120\% once affiliation is accounted for. The mechanism requires every participant to pay their bid regardless of outcome, which causes searchers to shade bids dramatically downward, and the aggregate shortfall more than offsets the contribution from losing bidders. All-pay formats also impose negative externalities on participants, discouraging entry and potentially reducing $n$ below the level assumed in the model.

    \item \textbf{Apply different mechanisms to different MEV types.} The competition parameters $(\rho, n)$ differ markedly across MEV categories. Sandwich attacks exhibit $n \approx 10$--$20$ and $\rho$ close to 1 (all searchers observe the same victim transaction), placing them near the non-monotone region of the revenue surface where high affiliation actually suppresses revenue. In this regime, the FPSBA or Dutch format may be preferred, since the English/SPSBA advantage is diluted by large $n$ and the collusion risk from repeat play is highest among sophisticated sandwich bots. By contrast, liquidations and naked arbitrage exhibit $n \approx 2$--$5$ and moderate $\rho$, the parameter region where the Linkage principle predicts the largest gains from open formats. A segmented approach, routing different MEV types to different mechanisms, can therefore outperform any single uniform format.
\end{enumerate}

\onecolumngrid
\clearpage
\appendix 
\section{Revenue Grids}
\renewcommand{\thetable}{A\arabic{table}}
\setcounter{table}{0}

\begin{table}[H]
\centering
\begin{tabular}{@{}ccccccccccc@{}}
\toprule
$n \backslash \rho$ & 0.0 & 0.1 & 0.2 & 0.3 & 0.4 & 0.5 & 0.6 & 0.7 & 0.8 & 0.9 \\
\midrule
2 & 5.41 & 6.55 & 8.04 & 9.86 & 12.39 & 15.24 & 18.87 & 23.93 & 30.56 & 40.59 \\
3 & 12.67 & 15.32 & 17.88 & 21.34 & 25.84 & 30.33 & 35.97 & 43.11 & 51.04 & 59.10 \\
4 & 21.05 & 24.51 & 29.05 & 33.20 & 39.43 & 45.31 & 52.70 & 60.41 & 67.43 & 75.35 \\
5 & 30.02 & 34.99 & 40.05 & 46.29 & 53.49 & 59.87 & 67.59 & 75.89 & 83.15 & 85.80 \\
6 & 39.61 & 45.57 & 51.32 & 58.58 & 66.13 & 74.19 & 81.28 & 89.21 & 94.15 & 97.71 \\
7 & 49.07 & 56.22 & 63.19 & 71.33 & 79.89 & 87.74 & 96.91 & 101.12 & 106.15 & 101.67 \\
8 & 59.47 & 66.83 & 75.28 & 83.51 & 92.96 & 101.07 & 110.90 & 115.11 & 119.43 & 113.37 \\
10 & 79.72 & 89.55 & 99.00 & 108.66 & 117.64 & 128.21 & 133.89 & 138.88 & 139.24 & 126.77 \\
12 & 100.91 & 111.23 & 122.20 & 133.23 & 142.00 & 152.07 & 156.81 & 164.28 & 155.66 & 137.43 \\
15 & 132.38 & 145.15 & 156.43 & 169.09 & 179.09 & 186.22 & 192.45 & 185.13 & 173.34 & 153.88 \\
20 & 185.55 & 199.95 & 214.48 & 225.99 & 236.26 & 238.33 & 235.08 & 232.37 & 211.43 & 170.86 \\
\bottomrule
\end{tabular}
\caption{\centering Expected revenue for English and SPSB auctions across $(n, \rho)$.}
\label{tab:appendix_1}
\end{table}

\begin{table}[H]
\centering
\begin{tabular}{@{}ccccccccccc@{}}
\toprule
$n \backslash \rho$ & 0.0 & 0.1 & 0.2 & 0.3 & 0.4 & 0.5 & 0.6 & 0.7 & 0.8 & 0.9 \\
\midrule
2 & 5.41 & 6.23 & 7.26 & 8.49 & 10.06 & 11.93 & 14.64 & 18.14 & 23.22 & 31.94 \\
3 & 12.71 & 14.45 & 16.39 & 18.72 & 21.73 & 25.03 & 29.31 & 34.67 & 41.33 & 50.03 \\
4 & 21.10 & 23.58 & 26.52 & 29.72 & 33.85 & 38.44 & 43.67 & 49.91 & 56.57 & 65.43 \\
5 & 30.06 & 33.33 & 37.02 & 41.21 & 46.10 & 51.10 & 57.37 & 64.34 & 70.75 & 75.70 \\
6 & 39.54 & 43.49 & 47.69 & 52.71 & 58.01 & 64.11 & 69.93 & 76.51 & 81.21 & 87.65 \\
7 & 49.32 & 53.87 & 58.85 & 64.15 & 70.18 & 76.64 & 83.14 & 87.85 & 93.57 & 93.19 \\
8 & 59.38 & 64.37 & 70.17 & 75.65 & 82.49 & 88.64 & 96.14 & 99.94 & 103.86 & 103.12 \\
10 & 79.89 & 86.08 & 92.57 & 98.91 & 105.57 & 112.48 & 119.30 & 121.82 & 123.67 & 116.08 \\
12 & 100.71 & 107.47 & 114.45 & 121.54 & 128.09 & 135.32 & 138.67 & 144.17 & 139.82 & 126.47 \\
15 & 132.67 & 140.06 & 147.51 & 154.63 & 161.93 & 166.36 & 171.75 & 167.09 & 157.97 & 143.00 \\
20 & 185.50 & 193.64 & 201.20 & 208.15 & 215.56 & 217.06 & 214.28 & 211.62 & 194.40 & 159.43 \\
\bottomrule
\end{tabular}
\caption{\centering Expected revenue for Dutch and FPSB auctions across $(n, \rho)$.}
\label{tab:appendix_2}
\end{table}

\begin{table}[H]
\centering
\begin{tabular}{@{}ccccccccccc@{}}
\toprule
$n \backslash \rho$ & 0.0 & 0.1 & 0.2 & 0.3 & 0.4 & 0.5 & 0.6 & 0.7 & 0.8 & 0.9 \\
\midrule
2 & 5.42 & 5.41 & 5.40 & 5.40 & 5.44 & 5.37 & 5.39 & 5.44 & 5.43 & 5.43 \\
3 & 12.71 & 12.82 & 12.66 & 12.65 & 12.79 & 12.78 & 12.73 & 12.70 & 12.71 & 12.66 \\
4 & 21.06 & 21.03 & 21.03 & 21.18 & 21.02 & 20.99 & 21.09 & 21.02 & 21.03 & 21.04 \\
5 & 29.85 & 30.11 & 30.33 & 30.08 & 30.05 & 30.04 & 30.03 & 30.22 & 29.96 & 30.04 \\
6 & 39.65 & 39.33 & 39.50 & 39.79 & 39.37 & 39.61 & 39.53 & 39.59 & 39.56 & 39.85 \\
7 & 49.10 & 49.47 & 49.46 & 49.44 & 49.16 & 49.24 & 49.35 & 49.33 & 49.41 & 49.15 \\
8 & 59.30 & 59.32 & 59.24 & 59.35 & 59.55 & 59.52 & 59.16 & 59.33 & 59.21 & 59.54 \\
10 & 79.57 & 79.79 & 79.67 & 80.08 & 79.88 & 79.93 & 79.90 & 79.84 & 79.96 & 79.98 \\
12 & 100.72 & 100.86 & 101.20 & 101.12 & 100.73 & 101.14 & 100.63 & 100.80 & 100.84 & 100.67 \\
15 & 132.26 & 131.92 & 132.64 & 132.83 & 132.43 & 132.53 & 132.75 & 131.78 & 131.89 & 132.80 \\
20 & 185.56 & 185.80 & 185.90 & 185.98 & 185.33 & 186.05 & 185.36 & 185.88 & 185.66 & 185.94 \\
\bottomrule
\end{tabular}
\caption{\centering Expected revenue for All-pay auction, which serves as the IPV setting benchmark. It shows constant values across $\rho$; any variation is caused by Monte Carlo noise.}
\label{tab:appendix_3}
\end{table}

\end{document}